\documentclass{article}
\usepackage[english]{babel}
\usepackage{subcaption}
\usepackage{bm}

\usepackage[a4paper,top=2cm,bottom=2cm,left=3cm,right=3cm,marginparwidth=1.75cm]{geometry}

\usepackage{float}
\RequirePackage{amsthm,amsmath,amsfonts,amssymb,adjustbox}
\RequirePackage[authoryear]{natbib}
\usepackage{authblk}
\usepackage[colorlinks=true, allcolors=blue]{hyperref}
\RequirePackage{graphicx,caption}
\graphicspath{{./images}} 
\usepackage{rotating}

\title{\textbf{A noisy-input generalised additive model for relative sea-level change along the Atlantic coast of North America}}
\author[*]{Maeve Upton}
\affil[*]{Hamilton Institute, Department of Mathematics \& Statistics, ICARUS, Maynooth University, Ireland}
\author[*]{Andrew Parnell}
\author[**]{Andrew Kemp}
\affil[**]{Department of Earth and Climate Sciences, Tufts University, U.S.A}
\author[$\ddagger$]{Erica Ashe}
\affil[$\ddagger$]{Department of Earth and Planetary Sciences, Rutgers University, U.S.A}
\author[$\dagger$]{Gerard McCarthy}
\affil[$\dagger$]{ICARUS, Maynooth University, Ireland}
\author[$\dagger \dagger$]{Niamh Cahill}
\affil[$\dagger \dagger$]{Department of Mathematics and Statistics, ICARUS, Maynooth University, Ireland}

\begin{document}
\maketitle
\begin{abstract}
We propose a Bayesian, noisy-input, spatial-temporal generalised additive model to examine regional relative sea-level (RSL) changes over time. The model provides probabilistic estimates of component drivers of regional RSL change via the combination of a univariate spline capturing a common regional signal over time, random slopes and intercepts capturing site-specific (local), long-term linear trends and a spatial-temporal spline capturing residual, non-linear, local variations. Proxy and instrumental records of RSL and corresponding measurement errors inform the model and a noisy-input method accounts for proxy temporal uncertainties. Results focus on the decomposition of RSL over the past 3000 years along the Atlantic coast of North America.
\end{abstract}

\section{Introduction}
The Intergovernmental Panel for Climate Change (IPCC) in 2021 reported with ``high confidence" that global mean rates of sea-level rise increased from approximately 1.3mm/yr between 1901 and 1971 to 3.7mm/yr between 2006 and 2018, with a further increase in rates predicted for the remainder of the 21st century \citep[p. 5]{IPCC2021summary}. In contextualising the socio-economic risk that this sea-level rise poses for coastal communities, it is necessary to place historic and predicted changes in a longer term (pre-anthropogenic) context and to recognise that local sea level can diverge sharply from the global average. 

\begin{figure}[H]
\centering
  \includegraphics[width=\textwidth]{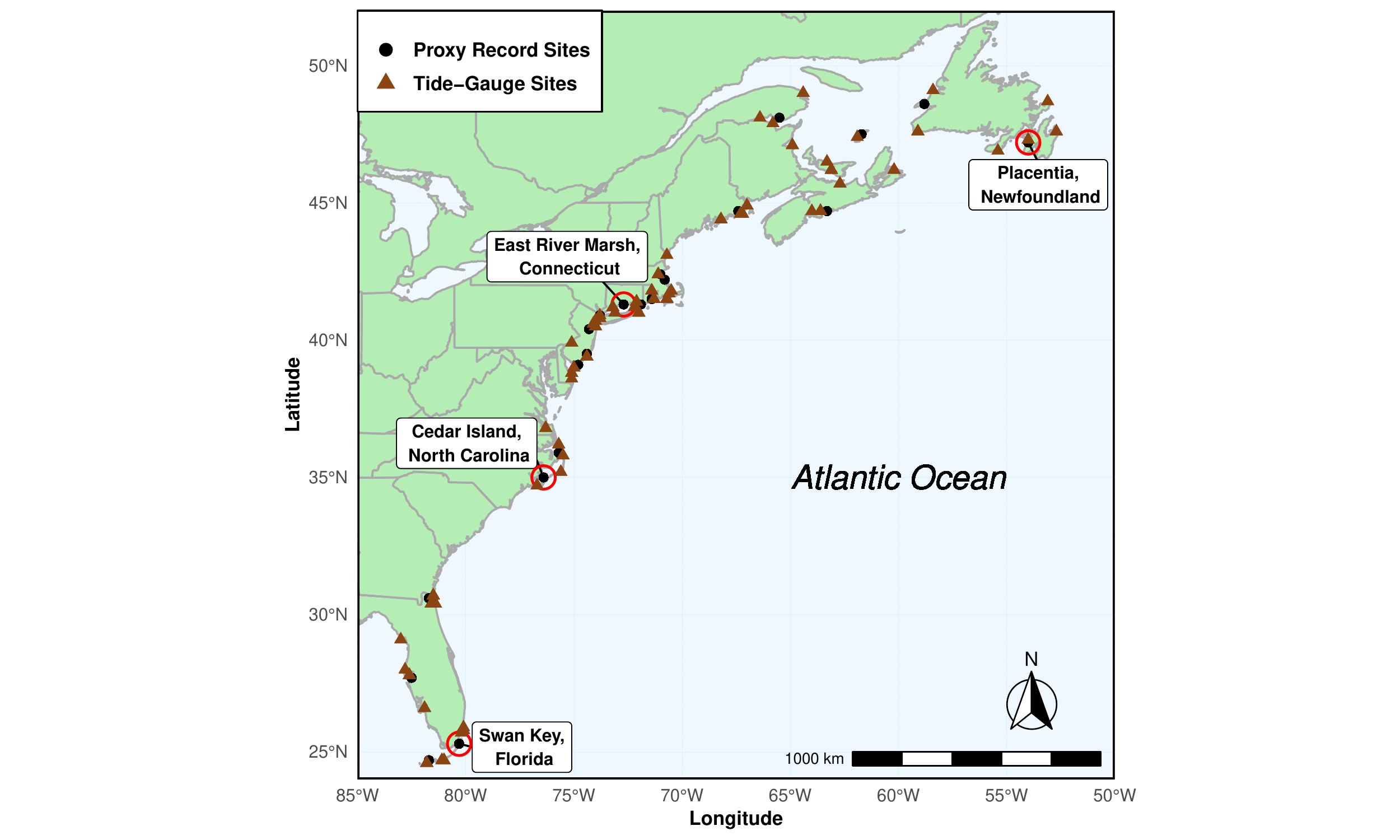}
  \caption{Location of the 66 tide gauge sites and 21 proxy record data sites along the Atlantic coast of North America with four proxy record sites chosen as case studies to present results of our model.}
    \label{fig:map}
\end{figure}

\begin{figure}[H]
  \centering
  \includegraphics[width=\textwidth]{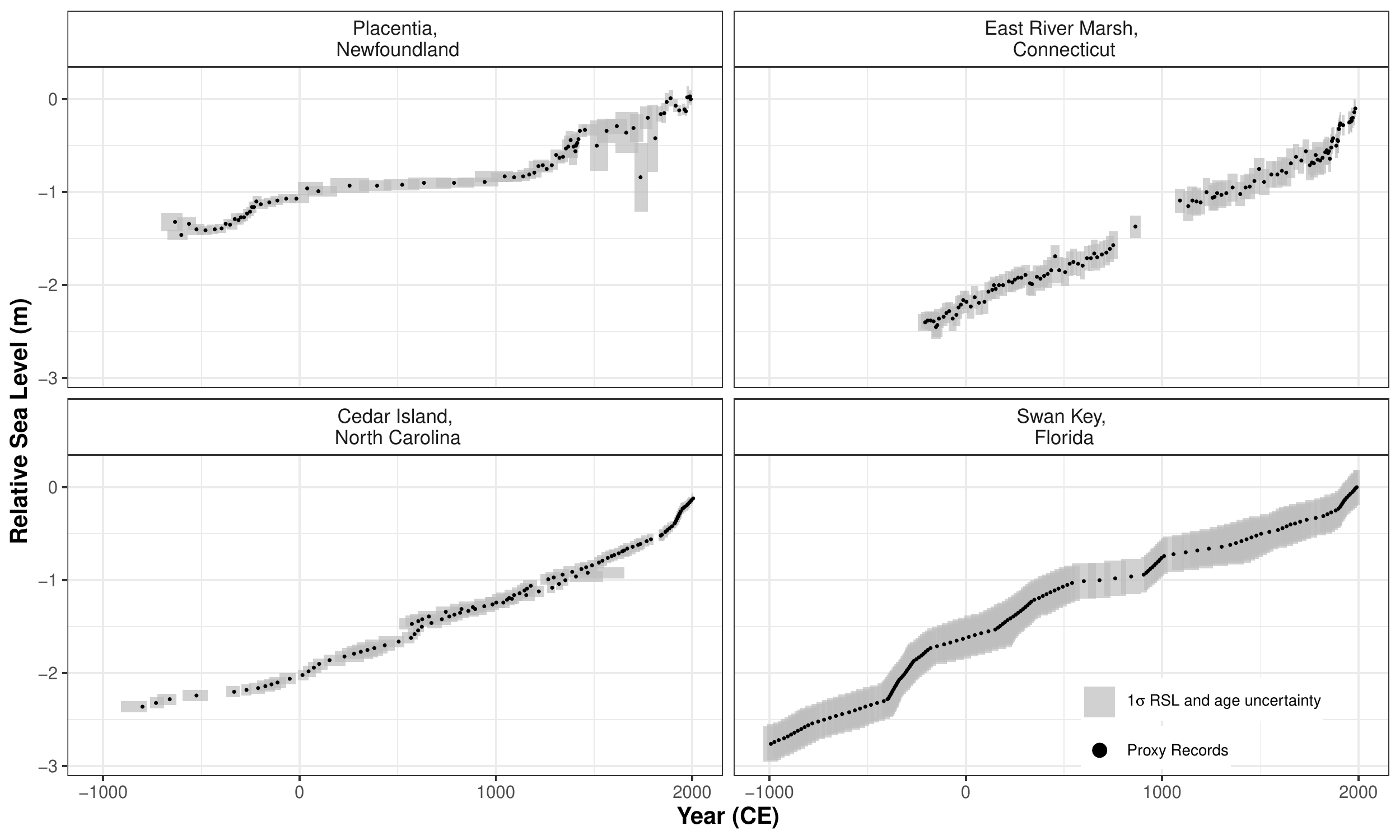}
\caption{Proxy records from four proxy sites along the Atlantic coast of North America used as illustrative case studies. The $y$-axis is relative sea level (RSL) in meters, where 0m is present sea level and negative values indicate RSL below present. Each proxy record observation consists of paired age and RSL estimate at the corresponding site. The black dot represents the midpoint of the proxy sea-level reconstruction and the grey boxes of 1 standard deviation represent vertical and horizontal (temporal) uncertainty.}
\label{fig:SL_data}
\end{figure}

Relative Sea Level (RSL) is the height of the ocean surface at any given location and time, measured relative to the adjacent land \citep{IPCC_2013_chp13}. Direct measurements of RSL (typically considered to be high accuracy and with low uncertainty) are made by a network of coastal tide gauges whose spatial distribution is highly uneven and whose temporal duration is typically limited to the past $\sim$ 100 years or less \citep{Church2011}. Understanding RSL before tide-gauge measurements began requires proxies  \citep[physical, biological, or chemical features with an "observable and systematic relationship to tidal elevation";][]{Horton2018MappingProbability} that are preserved in dated geological archives such as coastal sediment \citep[e.g.][]{Gehrels1994} or corals \citep[e.g.][]{Meltzner2017}. For the past 3,000 years (a period in Earth's history called the late Holocene), it is possible to generate near-continuous proxy RSL reconstructions which overlap tide-gauge measurements \citep{Kemp2013Sea-levelUSA}. The suite of late Holocene RSL proxy reconstructions is growing, but their global distribution is highly uneven \citep{Ashe2019}. However, the Atlantic coast of North America has a relatively large number of datasets (Figure \ref{fig:map}) generated from sediment that accumulated in salt-marsh \citep[e.g.][]{Kemp2018} and mangrove environments \citep[e.g.][]{Khan2022}. We therefore focus on this region to develop a new statistical model for quantifying patterns, rates, associated uncertainties and possible causes of late Holocene RSL change from a combination of proxy reconstructions and tide-gauge measurements concurrently.
A general discussion on how proxy records are developed is provided in Section \ref{data} and we point the reader in the direction of \citet{Shennan2015_Handbook} for a more detailed account of the methodologies employed by the paleo sea-level community. In this paper, we focus on analysing published data arising from proxy RSL reconstructions. The proxy records contain RSL estimates throughout time for different locations, specifically along the Atlantic coast of North America, and have associated bivariate uncertainties, i.e. uncertainty in time and vertical uncertainty in RSL (Figure \ref{fig:SL_data}).

Tide gauges and proxy records can only capture RSL, which is the net outcome of a complex combination of physical processes operating on characteristic temporal (years to millennia) and spatial (site-specific to global) scales. These physical processes often act simultaneously and serve to reinforce or mask one another; they can change both the height of the sea-surface and that of the land differently through time and across space \citep{Khan2022}. Consequently, RSL measurements can display a rich variety of spatio-temporal patterns. A principal goal of sea-level research is to interrogate these patterns to identify and quantify the contribution from specific physical processes including (but not limited to) the multi-millennial and regional response of the solid Earth to de-glaciation, decadal to centennial redistribution of ocean water by changing currents, and the recent (multi-decadal) acceleration of global average sea-level rise in response to a warming climate \citep{IPCC_2013_chp13}. This goal requires a means to decompose the site-specific RSL signal at each locality in the network into contributions at different temporal and spatial scales, while accounting for uncertainties in the underlying data.

The most widely used tool for decomposing late Holocene RSL is a model developed by \citet[][hereon K16]{Kopp2016} and its various extensions \citep{Kemp2018,Walker2021}. The K16 model decomposes RSL into three categories: (1) a non-linear signal common to all records in the dataset being analysed (termed global, irrespective of the geographic range of input data); (2) a regional signal characterized by a linear rate of change over the past $\sim$ 2000 years and (3) a local (site-specific) signal that operates in a non-linear fashion. Rather than representing specific physical processes, these categories serve to represent groups of processes that operate at similar spatial and temporal scales informed by the data. K16 employs Gaussian Process (GPs) for each component and, due to the associated computational burden which grows in proportion with the cube of the number of data points, relies on a maximum likelihood approach to estimate and fix model hyperparameters. These modelling decisions, that aim to reduce the computational burden of GPs, can impact uncertainty quantification \citep{Ashe2019}. In this paper we aim to propose an alternative method for estimating these complex, interdependent components, that improves uncertainty quantification whilst remaining computational feasible.

Our new spatio-temporal statistical approach to modelling RSL change uses Generalised Additive Models (GAMs). A GAM is a generalised linear model where ``the linear predictor depends linearly on a sum of smooth functions of the predictor variable" \citep[GAMs; ][p.~161]{wood_2017}. GAMs flexibly model non-linear relationships using smooth functions (most commonly splines) and can reduce computational complexity when compared with GPs (for example those used in K16) as they do not require large matrix inversions. We place our model in a Bayesian framework which allows for the estimation of parameters conditioned on the RSL data with full accounting for, and propagation of, uncertainty. Similar to K16, our model partitions the total RSL signal into components that characterize distinctive spatial and temporal scales, which (to varying degrees) are associated with specific physical processes. These components are: (1) a regional component, a non-linear signal common to all sites along the Atlantic coast of North America, and equivalent to the global term in K16; (2) a linear local component which contains unstructured random effects and is comparable the regional linear term in K16; and (3) a non-linear local component, which is site-specific and varies smoothly in space and time. Similar to K16, any variation not addressed by the model is captured by a residual term.

Since the data points at each site have a bivariate error structure, and the decomposition required involves differing structures, the simple application of default GAMs does not work in our case study. Previous methods, such as \citep{Cahill2015aStats}, provides guidance for how to model RSL with bivariate uncertainty. We follow K16 in accounting for the time error using the Noisy-Input uncertainty method of \cite{McHutchon2011}. This method inflates the residual variance by a corrective term to compensate for noisy-input measurements using a smooth process. Whilst the original paper uses the method exclusively for GPs we extend the approach to spline terms. The RSL error is captured via a standard measurement error term added to the residual variance.

The structure of our paper is as follows. Section \ref{data} addresses the proxy records and tide-gauge data used in our analysis. Section \ref{drivers_RSL} describes the main physical processes driving RSL changes and Section \ref{previous_model} discusses the previous modelling strategies employed by \citet{Kopp2016}. Section \ref{method} gives a detailed description of our statistical model with different splines representing each driver of RSL change and introduces the noisy-input method. The model validations are shown in Section \ref{modelval} and the results for different drivers of RSL change and their associated rates are presented in Section \ref{results}. Section \ref{discussion} provides concluding remarks for our approach for the Atlantic coast of North America.

\section{Data}\label{data}
We use a combination of instrumental data from tide gauges and proxy records. This section discusses the different sea-level data sets, including their collection methods and their associated uncertainties.

\subsection{Tide-Gauge Data}
Tide gauges are fixed to the land and regularly measure (for example hourly or to higher frequency) the height of the adjacent sea surface \citep{pugh_woodworth_2014}. For understanding RSL change, these observations are usually expressed as annual averages and held in the database maintained by the Permanent Service for Mean Sea Level \citep[PSMSL;][]{Woodworth2003, Holgate_PSMSL2013}. The $\sim$1500 stations in this global network display highly uneven distribution of data across space and through time and in addition individual records may have temporal gaps \citep{Church2011}.The earliest tide gauges records began in the late 17th or early 18th centuries in northwestern Europe \citep{Woppelmann2006}. Along the Atlantic coast of North America, the longest tide-gauge record in the PSMSL database is The Battery from New York City (since 1856 CE) \citep{Holgate_PSMSL2013}. Annual tide-gauge data from the PSMSL are treated as having fixed and known ages without uncertainty in elevation measurements \citep{Holgate_PSMSL2013}.

In our analysis, we use 66 tide-gauge sites along the Atlantic coast of North America (Figure \ref{fig:map}). Tide gauges meeting at least one of the following criteria were included in our analysis;
(1) record length exceeding 150 years; (2) the nearest tide gauge to proxy site; (3) within 1 degrees distance to a proxy site and longer than 20 years \citep{Kopp2016, Walker2021}. The addition of tide-gauge data supplements the long-term proxy records and provides additional insight into recent changes in RSL. Annual data for each tide gauge were downloaded from the PSMSL and expressed in meters relative to the average over 2000-2018 CE. This time window captures variability resulting from the 19-year cycle in astronomical tides \citep{pugh_woodworth_2014b} and serves to make proxy and tide-gauge data comparable since the sediment cores used to develop proxy reconstructions were recovered since $\sim$ 2000 CE. In addition, we further average tide-gauge data by decade to increase comparability with proxy reconstructions that are developed from 1 cm thick slices of core sediment which accumulated over a period of several years (depending on sedimentation rate) and are therefore inherently time averaged. At this step we include an uncertainty in the tide-gauge data (± 5 years for age and ± 1$\sigma$ for RSL). See Appendix \ref{appendix_full_data} for additional information.

\subsection{Proxy Records}
Proxy-based reconstructions provide estimates of pre-anthropogenic RSL \citep{Kemp2013Sea-levelUSA}. On the Atlantic coast of North America, these near-continuous proxy-based reconstructions are generated using buried sequences of salt-marsh \citep[at mid to high latitudes;][]{Gehrels2020} or mangrove \citep[low latitudes;][]{Khan2022} sediment. Samples of this sediment are recovered in a core (a column of sediment extracted from the ground, where the oldest material is at the bottom and the youngest material is at the top) and interrogated in subsequent laboratory analysis to determine the age of the sample and the tidal elevation (height above a tide level) at which it accumulated \citep{Horton2006}. 

A history of sediment accumulation provides estimates of sample ages by directly dating a subset of depths in the sediment core, typically using radiocarbon measurements \citep{Torn2015SL_handbook}. In addition, the shallowest (i.e. most recent) part of the core can be dated by recognising historic pollution and land use changes of known age in down-core profiles of elemental abundance, isotopic activity and isotopic ratios \citep{Marshall2015_SLhandbook}. These directly dated levels in the core are the input (i.e. age of sediment sample) for a statistical age-depth model (e.g. the Bchron \citep{parnell2008}, Bacon \citep{Blaauw2011BACON}, or Rplum \citep{Aquino2018_Rplum} packages in R). These age-depth models (irrespective of their specific similarities and differences) estimate the age of every 1 cm thick sediment sample in the core with uncertainty. Comparisons indicate that sediment accumulation histories have little dependence on the specific age-depth model used \citep{WRIGHT2017}. 

 A sea-level proxy is required to reconstruct RSL. A sea-level proxy is any physical, biological or chemical feature with an observable and systematic relationship to tidal elevation \citep{Shennan2015_Handbook}. Salt marshes and mangrove environments are vegetated by distinctive plant communities that are adapted to inundation by salt water, resulting in distinct and narrow elevation ranges \citep{Redfield1972}. This distribution makes salt-marsh vegetation a valuable sea-level proxy. Through reasoning by analogy, the observable distribution of plants in modern salt marshes enables interpretation of their analogous counterparts preserved in core material \citep{Kemp2015_SLhandbook}. In this way, the paleo-marsh elevation (elevation with respect to tidal elevation at the time of formation) is reconstructed. Another sea-level proxy preserved in salt-marsh sediment is the remains of micro-fossils (e.g., foraminifera) that form distinctive assemblages with a strong relationship to elevation \citep{Edwards2015_SLhandbook}. When using micro-fossils to reconstruct RSL a transfer function is required which relates the abundance of specific micro-fossil families to tidal elevation using a dataset that is representative to the modern environment \citep{Kemp2015_SLhandbook}. There are various transfer functions available using Frequentist \citep{Sachs1977, Horton2006, Kemp2011} and Bayesian approaches \citep{Cahill2016}, which all estimate paleo-marsh elevation with uncertainty. The age of each core sample with a paleo-marsh elevation reconstruction is provided by the age-depth model. Resulting in a single proxy RSL record comprised of stratigraphically-ordered data points of age (with 1 sigma uncertainty) and RSL (with 1 sigma uncertainty).

We analyze 21 RSL proxy records (totaling 1731 data points) located on the Atlantic coast of North America from the Florida Keys, USA to Newfoundland, Canada (Figure \ref{fig:map}). There are 66 tide-gauge records that meet our criteria for inclusion (Figure \ref{fig:map}). The spatial scope of our analysis is restricted to this coastline because it has (by a considerable margin) the greatest concentration of available records. Results presented in this paper are generated from all the proxy and tide-gauge records, yet we present four of these sites (Placentia Newfoundland Canada, East River Marsh Connecticut USA, Cedar Island North Carolina USA, and Swan Key Florida) as illustrative case studies throughout the remainder of the manuscript. The four sites were selected to provide diversity of location and therefore the processes causing RSL change during the past $\sim$ 3000 years.

\section{Drivers of RSL Change}\label{drivers_RSL}
Spatio-temporal models recognise sea-level variability characteristic of different spatial and temporal scales rather than from specific processes. Each component estimated in the model may capture several contributing processes depending on the location and time interval under examination. These processes may act simultaneously and in directions that mask or exaggerate contributions from other drivers.

Transfer of mass between land-based ice and the ocean drives RSL change. Ice melt/growth returns or removes mass to the ocean as liquid water which causes a rise/fall in global mean sea level (this process is termed barystatic) \citep{Gregory2019ConceptsGlobal}. This contribution varies in magnitude across timescales, but is common to all locations. In addition, changes in global temperatures alter the density of ocean water resulting in a sea-level change (rise/fall when water warms/cools becoming less/more dense); this process is known as a thermosteric contribution \citep{Grinsted2015}. The global term in the K16 model and the regional component in our model attempts to capture influences from these processes.

Along the Atlantic coast of North America, the principal driver of RSL change during the pre-industrial late Holocene is glacial isostatic adjustment (GIA) \citep{Roy2015}. GIA is the response of the Earth, the gravitational field, and the ocean to the growth or decay of ice sheets \citep{Whitehouse2018}. GIA can be reasonably approximated as a linear contribution through time on this relatively short timescale, but with considerably variability along the coast \citep{Engelhart2009}. There are a family of physical models known as Earth-ice models which use a representation of the physical Earth structure (such as lithospheric thickness and properties such as mantle viscosity) to predict changes in GIA that occur through loading and unloading of ice, and provide estimates of GIA rates. One such example of an Earth-ice physical model is the ICE5G VM2-90 \citep{Peltier2004}. It is important to recognise that other processes (e.g., tectonically-driven vertical land motion) can mimic the linear trend of GIA. However, along the passive margin of the Atlantic coast of North America these non-GIA drivers are likely modest in magnitude \citep{Kopp2015}. As a result, the linear local component in our model and the linear regional term in the K16 may capture contributions from processes other than GIA that drive RSL changes.

There are processes with a spatially-coherent structure where the signal is shared by some but not all sites \citep{Stammer2013}. One such process that can cause RSL to vary on decadal to multi-century timescales is the redistribution of existing ocean mass by shifts in prevailing patterns and strength of atmospheric and oceanic circulation (termed dynamic sea-level change) \citep{Gregory2019ConceptsGlobal}. Dynamic sea level varies by site, but the magnitude of the difference from one site to the next is too small to be detected using proxy data due to the resolution. Some processes (e.g. sediment compaction: which can impact the height of the solid Earth surface with changes in sediment volumes for each site \citep{Horton2018MappingProbability}) can drive RSL changes that are site-specific. Consequently, contributions from these processes lack spatial coherence and display an unpredictable spatial structure. Therefore, site-specific RSL changes can vary markedly across closely-spaced sites. More often than not, RSL proxy reconstructions are not generated with the goal of understanding site-specific processes \citep{Walker2021}. It remains important to quantify this component as a means to distill the contribution from processes acting at larger spatial scales. In our model the structured (common to some, but not all sites) and unstructured (unique to one site) RSL variability on century timescales is captured by the non-linear, local component.

\section{Previous Statistical Models for RSL change} \label{previous_model}
In this section, we review previous work on modelling RSL change, focusing in particular on K16. The model was further extended in \cite{Kemp2018} and \cite{Walker2021}, here, we focus on the simpler K16 model. We first review the structure of this model, which decomposes RSL into component parts before discussing how the model might be fitted to the data and the potential influence of optimising hyperparameters using maximum likelihood. K16 forms the basis upon which we build our new approach in Section \ref{method}.

The RSL measurements are recorded in units of height; with meters used by default. In cases where the scale of the change is relatively small we use cm or mm instead for some plots and discussion in the text. We write $y_{ij} = y(\mathbf{x}_j, t_{ij})$ for the RSL height at location $\mathbf{x}_j$ (latitude and longitude) and time $t_{ij}$. These observations arise from the proxy records and tide gauges with $j$ indexing the data site and $i$ the observation. For the resolution of the data the time is expressed in years CE. The K16 model can be written as:
\begin{equation}\label{eq:K16_1}
y_{ij} = f(\mathbf{x}_j,t_{ij}) + w(\mathbf{x}_j,t_{ij}) + y_0(\mathbf{x}_j) + \epsilon_{ij}^y
\end{equation}
where $f$ is the full RSL spatio-temporal field, $w$ is a white noise process representing sub-decadal trends unexplained by the data due to resolution of data, $y_0$ is a site-specific spatially variable vertical offset, and $\epsilon_{ij}^y$ is residual error. In K16 all the structured terms above are given Gaussian Process prior distributions with stationary covariance functions. 

A key complication is that the times $t_{ij}$ associated with the proxy records are observed with uncertainty. Thus the observed values $\tilde{t}_{ij}$ have measurement error, defined as:
\begin{equation}\label{eq:K16_2}
\tilde{t}_{ij} = t_{ij} + \epsilon_{ij}^t.
\end{equation}
Usually $\epsilon_{ij}^t$ is assumed iid normally distributed with known variance, though in reality the age-depth model through which the ages are estimated often provides skewed distributions. A previous attempt at resolving this issue can be found in \citet{Cahill2015aStats}, though across large, multi-site datasets the imposition of this assumption is believed to have minor effects on the outcome of the model \citep[as shown in][]{Parnell2015_Handbook}. 

\subsection{Decomposing the RSL field \textit{f}}

For the RSL process defined above as $f$, K16 use a spatio-temporal empirical Bayesian hierarchical model to partition the influence of the components into global, regional and local scales. The fields that make up $f$ are, as above, given stationary GP priors that can vary in time and space as controlled by the covariance functions \citep{Ashe2019}. The standard decomposition of $f$ is written:
    \begin{equation}\label{eq:K16_3}
        f(\mathbf{x}_j,t_{ij}) = c(t_{ij}) + g(\mathbf{x}_j)(t_{ij} - t_0) + l(\mathbf{x}_j,t_{ij})
    \end{equation}
where $c(t_{ij})$ is term the global term, the temporal non-linear signal common across all sites, designed to capture changes such as barystatic sea level rise and thermosteric changes. $g(\mathbf{x}_j)$ is a spatially varying term that captures slower processes such as long-term land level change (GIA) and vertical land motion driven by plate tectonics. The $g$ term is multiplied by time $t$ differenced from a reference point $t_0$ to form temporally linear field. Unlike the other components in K16, $g$ is given a univariate normal prior distribution with the mean centred on the value obtained from a Earth-ice physical model \citep[ICE5G VM2-90;][]{Peltier2004} which estimates the GIA rate. $l(\mathbf{x_j}, t_{ij})$ is the local spatio-temporal field that describes factors such as dynamic sea level change, sediment compaction and tidal regimes. These terms are explained in more detail in K16.

Without strong prior information, it is difficult to separate out the magnitudes of the components. Thus in K16, the hyperparameters are first obtained by maximising the likelihood of the model conditioned on the observations but constrained to two timescale hyperparameters for the non-linear terms. The model is then re-fitted using these hyperparameters to estimate the components of the fields in an empirically Bayesian framework. In our approach, we aim to avoid the empirical Bayesian approach of fixing hyperparameters by placing informed priors on the model components. However, model shortcuts are unavoidable due to the complexity of the decomposition and the innate confounding of many of the key terms. 

\section{A new approach based on Generalised Additive Models}\label{method}

In this section we outline a new approach to evaluating the different drivers of spatio-temporal RSL using proxy records and tide gauge data. With careful choices of the prior distributions of the hyperparameters, we aim to recover the components of RSL change through the standard tools of Bayesian inference. Subsequently, we estimate of rates of RSL change at sites along the Atlantic coast of North America. We build our model inspired by the standard decomposition of the RSL field $f$ as described above. Our approach contains four main differences compared to the \citet{Kopp2016} (and subsequent) models:
\begin{enumerate}
    \item We focus on the high quality sites along North America's Atlantic coast and aim to produce a regional RSL curve. Thus we avoid making statements about global sea-level change.
    \item We use splines instead of GPs to avoid the computationally challenging inversion of the GP covariance matrices. The model, at its simplest, thus falls under the standard generalised additive modelling paradigm.
    \item We fit the model in two stages to maximise the regional variability which would otherwise be confounded with the local structure. This allows us to perform a more complete posterior analysis of the model hyperparameters which might otherwise been fixed in K16.
    \item We remove the spatial structure on the linear effect $g$ in K16 and replace it with a univariate random effect on the slope. For the proxy records we use a prior mean for the slope that is informed by the data before 1800 CE \citep[i.e. the pre-industrial time period;][]{2K_2019}. This change is helpful because we have found the estimated values of the GIA rate from the Earth - ice physical model \citep[e.g.][]{Peltier2004,Peltier2014,Caron2018} do not match the observed data well for the proxy record time period. For the tide gauge records the prior mean of the slope is taken from a physical Earth-ice model \citep[ICE5G VM2-90;][]{Peltier2004} with uncertainty taken from \citet{Engelhart2009}.  We refer back to this modelling choice in Section \ref{discussion}. 
\end{enumerate}

Below we outline the full posterior distribution of the model to highlight the assumed conditional independences, then outline each term and its structure. The temporal uncertainty in the data causes difficulties in fitting the model in one step, and we resort to \citet{McHutchon2011}'s noisy-input method to account for this uncertainty. We then discuss the prior distributions assumed for the hyperparameters, and the computational details of our model. In Section \ref{results}, we showcase the successful implementation of our model.

\subsection{Model Notation}
\noindent We now provide a full outline of our notation for reference:
\begin{itemize}
    \item $y_{ij}$ is an RSL observation in meters with $i = 1,...n_j$ observations at site $j$ with $j = 1,..., m$ sites. We vectorise the full set of observations as  $\mathbf{y}$ and the observations for each site as $\mathbf{y}_j$. 
    \item $t_{ij}$ are the ages of each RSL observation, indexed and vectorised as above. We represent age in years of the Common Era (CE).
    \item $\mathbf{x}_j$ is the 2-vector of a latitude and longitude pair for each site $j$.
    \item $z_{\mathbf{x}}$ is an index vector for the data sites that converts each site into a label. Thus $z_{\mathbf{x}_j} = j$.
    \item $f(\mathbf{x}_j,t_{ij}) = f_{ij}$ is the mean sea-level process at site $j$ and time $t_{ij}$. We write $f(\mathbf{x},t)$ as the mean process for a generic location and time, and continue with this notation below for brevity
    \item $r(t)$ is the regional component at time $t$. 
    \item $l(\mathbf{x},t)$ is the non-linear local component at location $x$ at time $t$.
    \item $g(z_{\mathbf{x}})$ is the linear local component at location $x$.
    \item $h(z_{\mathbf{x}})$ is a site-specific vertical offset component at location $x$.
    \item $b_{r}(t)$ and $b_l(\mathbf{x},t)$ are sets of known b-spline basis functions corresponding to the regional and local components respectively.
    \item $\bm{m}^{g}$ and $\bm{s}^{g}$ are the mean and standard deviation parameters respectively for the linear local correction component. These are site specific and so each is a vector of length $m$.
    \item $\bm{\beta}^r, \bm{\beta}^l$ are the spline regression coefficient vectors of the regional and local components respectively. $\bm{\beta}^r$ is of length $k_r$ and $\bm{\beta}^l$ is of length $k_l$ where $k_r$ and $k_l$ are the number of knots associated with each term.
    \item $\bm{\beta}^g,\bm{\beta}^{h}$ are parameter vectors, each of length $m$, containing the  random effect coefficients for each site.
    \item $\sigma_r$ and $\sigma_l$ are the smoothness parameters associated with the regional and local spline terms respectively. 
    \item $\sigma_{h}$ is the standard deviation of the site-specific offset.
    \item $s_{y_{ij}}$ is the known standard deviation of the RSL data point $ij$.
    \item $s_{t_{ij}}$ is the known standard deviation of the age of data point $ij$. 
    \item $\sigma$ is a residual standard deviation parameter to capture any remaining variability in $y$.
\end{itemize}

\subsection{Posterior Distribution}
The joint posterior distribution of our Bayesian hierarchical model is shown below:
\begin{multline}
    \underbrace{p(\sigma^2, \bm{\beta}^{r}, \bm{\beta}^{l},\bm{\beta}^{g},\bm{\beta}^{h}, \sigma_r^2, \sigma_l^2, \sigma^2_{h}| \mathbf{y}, \bm{b}_{r}, \bm{b}_{l}, m_g, s_g^2, \mathbf{s}_y^2, \mathbf{s}_t^2)}_\text{posterior} \propto
    \underbrace{p(\mathbf{y}|\bm{f}, \sigma^2, \mathbf{s}_y^2, \mathbf{s}_t^2)}_\text{likelihood} \times \\ \underbrace{p(\bm{\beta}^{r}| \sigma_{r}^2)}_\text{prior on regional parameters} \times  \underbrace{p(\sigma_{r}^2)}_\text{prior on regional smoothness parameter} \\ \times
    \underbrace{p(\bm{\beta}^{l}|\sigma_{l}^2)}_\text{prior on non-linear local parameters} \times \underbrace{p(\sigma_{l}^2)}_\text{prior on non-linear local smoothness parameter} \times
    \underbrace{p(\bm{\beta}^{g}|\bm{m}_{g},\bm{s}^2_{g})}_\text{prior on linear local parameters}\\ \times
    \underbrace{p(\bm{\beta}^{h}|\sigma^2_{h})}_\text{prior on site-specific vertical offset parameters} \times \underbrace{p(\sigma^2_{h})}_\text{prior on variance site-specific vertical offset parameters} \times
    \underbrace{p(\sigma^2)}_\text{prior on error variance}
\end{multline}
The likelihood $p(\mathbf{y}|\bm{f}, \sigma^2, \mathbf{s}^2_y, \mathbf{s}^2_t)$ can be deconstructed thus:
\begin{equation}
         p(\mathbf{y}|\bm{f}, \sigma^2, \mathbf{s}_y^2, \mathbf{s}_t^2) = \prod_{j=1}^m \prod_{i = 1}^{n_j} \mathcal{N}(y_{ij}|f_{ij}, \sigma^2 + s_{y_{ij}}^2 + s_{t_{ij}}^2)
\end{equation}

\subsection{A fully specified generalised additive model for decomposing the RSL field}\label{subsec:model_desc}

Our version of the decomposition of the mean sea level field can be written as:\begin{equation}\label{eq:NIGAM_process}
       f(\mathbf{x},t) =  r(t)+ g(z_\mathbf{x}) + h(z_\mathbf{x}) + l(\mathbf{x},t)+ \mathbf{\epsilon_y}
    \end{equation}
All terms are as defined above: $r(t)$ is the regional component. $g(z_{\mathbf{x}})$ is the linear local component represented by a random effect with $z_{\mathbf{x}}$ representing each data site. $h(z_\mathbf{x})$ is the spatial vertical offset for each data site. $l(\mathbf{x},t)$  is the non-linear local component.  
We represent $r(t)$ using a spline:
\begin{equation}\label{eq:regional}
r(t) = \sum^{k_r}_{s=1} b_{r_s}(t)\beta^{r}_s
\end{equation}
where $\beta^{r}_s$ is the $s^{th}$ spline coefficient, $k_r$ is the number of knots and $b_{r_s}(t)$ is the $s^{th}$ spline basis function at time $t$.

The linear local component, $g(z_{\mathbf{x}})$, is an unstructured random effect for each site which is formulated as:
\begin{equation}\label{eq:linear_loc}
g(z_{\mathbf{x}_j}) = \beta^{g}_jt
\end{equation}
where $\beta^{g}_j$ is a slope parameter specific for each site $j$.  This specification is in contrast to K16 where the linear effect, $g$, varies smoothly in space and is informed through the prior by GIA model-derived values. We found such a restriction to adversely affect model performance due the lack of agreement between the data and the provided GIA values, and the wide variation in values between proximal sites \citep{Engelhart2009}. 

The site-specific vertical offset $h$ is a random effect used to capture vertical shifts associated with measurement  variability between sites and is formulated as:
\begin{equation}\label{eq:site_offset}
h(z_{\mathbf{x}_j}) = \beta^{h}_j
 \end{equation}
where $\beta^{h}_j$ contains the random effect coefficients for site $j$.

The non-linear local component $l(\mathbf{x}, t)$ is described with a spatio-temporal spline function formulated by:
\begin{equation}\label{eq:nonlinear_loc}
l(\mathbf{x}, t) = \sum_{s=1}^{k_l} b_{l_s}(\mathbf{x},t) \beta^{l}_s
\end{equation}
where $ \beta^{l}_s$ is the $s^{th}$ spline coefficient, $k_l$ is the number of knots and $b_{l_s}(\mathbf{x},t)$ is the $s^{th}$ spline basis function at time $t$ and location $\mathbf{x}$.  

We use B-splines \citep{deBoor1978} for both the regional and local terms. Our B-splines are constructed as piece-wise polynomials which join together at equidistant knots such that the first derivatives are equal \citep{eilers_1996}. For the regional term we use cubic B-splines as we are interested in the behaviour of the first derivatives. We can simply calculate these by differentiating the cubic B-splines and multiplying with the posterior spline parameters to provide a posterior distribution for the derivative. However for the non-linear local component, which requires a tensor product to capture the variability over time and space (represented with longitude and latitude) so that the individual covariates are combined product-wise \citep{Wood2006a}. We use a simpler and faster quadratic polynomial basis for the non-linear local component. Many other basis function types and options are available \citep[see, e.g.][]{dierckx1995curve, wood_2017} but we believe our approach balances both parsimony and computational efficiency for our application area. 

\subsection{Noisy-Input Uncertainty Method}\label{NoisyGam}
Our data is corrupted with measurement error in the RSL values and that arising from the temporal uncertainty associated with radiocarbon dating the fossil layers of sediment. \cite{McHutchon2011} describe an assumption for GPs which avoids the need for complex errors-in-variables models \citep[e.g.][]{Dey2000,Cahill2015aStats} and instead adds an extra measurement uncertainty on the response variable. We adapt this noisy-input (NI) approach for our RSL GAM which we now term an NI-GAM.

The response variable $y$ is assumed to be a noisy measurement with the true output given as $\tilde{y}$:
\begin{equation}\label{eq:NI_1}
 y = \tilde{y} + \epsilon_y
\end{equation}
where the error term is given by $\epsilon_y \sim \mathbb{N}(0, s_y^2)$ with $s_y$ being the known measurement standard deviation of the RSL data. Similarly, for the input measurements, $t$ is assumed to be a noisy estimate of the true time value $\tilde{t}$:
\begin{equation}\label{eq:NI_2}
 t = \tilde{t} + \epsilon_t
\end{equation}
with the error term given by $\epsilon_t \sim \mathbb{N}(0, \tilde{s}_t^2)$ where $\tilde{s}_t$ is the known standard deviation of the date obtained from the age-depth model (described in Section \ref{data}).
As a result, a function for the response variable is formed in the following way:
\begin{equation}\label{eq:NI_3}
y = f(\mathbf{x},\tilde{t} + \epsilon_t) + \epsilon_y
\end{equation}
Following \cite{McHutchon2011} we can use a Taylor expansion about the latent state $\tilde{t}$ to obtain:
\begin{equation}\label{eq:NI_4}
f(\mathbf{x},\tilde{t} + \epsilon_t) = f(\mathbf{x},\tilde{t}) + \epsilon_t^T\frac{\partial f(\mathbf{x},{\tilde{t}})}{\partial \tilde{t}} + \cdots \approx f(\mathbf{x},t) + \epsilon_t^T\frac{\partial f{(\mathbf{x},t)}}{\partial t}
\end{equation}
Thus the error in $t$ can be approximated by an increase in the measurement error proportional to the derivative of $f$. \cite{McHutchon2011} calculate the derivative of the mean of the GP function, given as vector $\partial_{\bar{f}}$ for the first order case and $\Delta_{\bar{f}}$ for a $D$-dimensional matrix.

Analogously for our NI-GAM setting, the first order terms are expanded to form a linear model with input noise:
\begin{equation}\label{eq:NI_5}
y = f(\mathbf{x},t) + \epsilon_t^T\partial _{\bar{f}} + \epsilon_y
\end{equation}
The derivative of the posterior mean for $f$ is obtained using a two-step method. First the model is fitted ignoring the input uncertainty and then the slope of the posterior mean is calculated. From this, a corrective variance term can be calculated, which we write as $s_t^2$. We use this as an additional model error term in our subsequent full model fit.

Intuitively, the input noise impacts the gradient of the function mapping input to output and the input noise variance is related to the output by the square of the posterior mean function’s gradient \citep{McHutchon2011}. As a result, the corrupted input measurements influence a rapidly changing output value more than an output value that remains constant. The advantage of this method is that the noise remains the same whether the measurement is considered an input or output measurement, and so all the data informs the input noise variance ensuring the output dimensions are met, reducing the chance of over-fitting.

\subsection{Prior distributions}\label{priors}
Within the process level each component is given a prior distribution. Our prior for the spline coefficients of the regional component, $\beta^r_s$ is:
\begin{equation}\label{eq:prior_reg}
    \beta^r_s \sim \mathbb{N}(0,\sigma^2_{r})
\end{equation}
where $\sigma_{r}$ is the standard deviation of the spline coefficient and fundamentally controls the smoothness of the model fit.

Our prior for the linear local component for the proxy records is:
\begin{equation}\label{eq:prior_linear}
    \beta^{g}_j \sim \mathbb{N}(m_{g_j}, s^2_{g_j})
\end{equation}
where $m_{g_{j}}$ and $s^2_{g_{j}}$ are the empirically estimated rate and associated variance for the data set obtained from the time period prior to 1800 CE \citep{2K_2019}. For the tide-gauge records, we obtain $m_{g_{j}}$ from a physical model \citep[ICE5G VM2-90;][]{Peltier2004} and $s^2_{g_{j}}$ from previous studies \citep{Engelhart2009}.

Our prior distribution for the site-specific vertical offset is:
\begin{equation}\label{eq:prior_offset}
    \beta^{h}_j \sim \mathbb{N}(0, \sigma^2_{h})
\end{equation}
where $\sigma^2_{h}$ is the variance of the random intercept across data sites.

Our prior on the spline coefficient for the non-linear local component is given as:
\begin{equation}\label{eq:prior_local}
    \beta^l_s \sim \mathbb{N}(0,\sigma_{l}^2)
\end{equation}
where $\sigma^2_{l}$ is the variance of the spline coefficients over space and time. This parameter fundamentally controls the smoothness of the local non-linear effect.

The remaining hyperparameters of the model include $\sigma^2_{r}$, $\sigma^2_{h}$ and $\sigma^2_{l}$.
The standard deviation parameter $\sigma_{h}$ represents the variability in the site specific vertical shift after taking account of the local linear trend. As it is measured in meters it is more interpretable in a physical context and so we place an informative prior here. The vertical shifts can be quite variable with some sites sitting many meters above or below others. From revisiting the publications associated with our data (see Appendix \ref{appendix_full_data} for the full list), shifted values spanning more than 5m seem unlikely. As a result, we specify the standard deviation to have a Cauchy distribution with mode 2.5m but with a wide scale of a further 2m \citep{gelman2006prior}. For the variability of the spline coefficients across the knots, i.e. $\sigma^2_{r}$ and  $\sigma^2_{l}$, we expect considerably smaller variation but we have less information, thus we use a truncated Cauchy distribution centred on zero and with scale value 1.

\subsection{Model fitting and computational details}
In previous sections, we have described our Bayesian hierarchical model structure using GAMs and the manner in which we account for uncertainty. In this section, we address how to fit our NI-GAM model and the decisions that influenced our model fitting strategy. We are constrained because we have to fit the model twice as part of the noisy-input uncertainty method, described in Section \ref{NoisyGam}. We also found that a single model fit yielded poor convergence due to the unavoidable confounding between the regional, vertical shift, linear, and non-linear local terms. Thus, we use the two-stage NI process to our advantage by fitting a slightly reduced model at the first stage to estimate the regional term, and using the posterior as strong prior information in the second stage to provide the estimate of the non-linear local term. Our approach has some similarities to that of cut feedback or modularised Bayesian models \citep{Plummer2015}, but we do not explore these avenues further here. 

The two steps of our model fit are:
\begin{enumerate}
    \item We first fit a simplified version of our process level model where we replace $f(\mathbf{x},t)$ with $f^*(\mathbf{x},t)$, defined as:
\begin{equation}
    f^*(\mathbf{x},t) = r(t) + g(z_{\mathbf{x}}) + h(z_\mathbf{x})
\end{equation}
This removes the non-linear local component and so avoids the confounding issue. From this model fit we calculate the first derivative of the posterior mean of $f^*(\mathbf{x},t)$. The resulting slope estimate for each observation provides a corrective variance term, $s_{t_{ij}}^2 = \tilde{s}_{t_{ij}}^2\partial _{\bar{f_{ij}^*}}^2$.  This term is added to the other model error variances for the fit in the second stage.

\item In the second step we fit the complete process model as defined in Section \ref{NoisyGam}. The only changes being: (1) the addition of the new noisy-input measurement variance term; (2) the prior distribution on the regional spline terms now being $\beta^r_s \sim \mathbb{N} (m^r_s,(s^{r}_s)^2)$ where $m^r_s$ and $s^{r}_s$ are estimated in the first model run; and (3) the prior distribution on the vertical offset term being $\beta^{h}_s \sim \mathbb{N}(m^h_s,(s^h_s)^2)$ where as above $m^h_s$ and $s^{h}_s$ are estimated in the first model run. 
\end{enumerate}

In effect the second model fitting stage simply becomes a means by which the full error uncertainty is accounted for and the residuals are decomposed into a pure error and a non-linear local space-time effect. Whilst all our subsequent results are presented based on the second model fit, this stage is essentially only useful for providing interpretation of the model error and the degree to which local factors drive deviations from the main regional effect. 

At each stage our models are written using the Just Another Gibbs Sample \cite[JAGS][]{plummer2003jags} software, which in turn is based on the on \citet{Spiegel2002}. The JAGS language uses Gibbs sampling and the Markov Chain Monte Carlo (MCMC) algorithm to draw samples from the posterior distribution of the unknown parameters. We implement our approach using the \texttt{rjags} package in R \citep{plummer2016rjags}. 
For our models, we used 2000 iterations with a burn-in value of 1000, thinning at 5 and 2 chains. Convergence diagnostics for the parameters are investigated using the coda package \citep{plummer2006coda} and the ShinyStan app which provides an interactive visualization tool for investigating model convergence \citep{Gabry2017shinystan}. All convergence diagnostics were checked and ensured to be satisfactory before the model results were interpreted. The code and data for our model can be found \href{https://github.com/maeveupton/NI-GAM.git}{here}.

\section{Model Validation}\label{modelval}

We examine the validity of our model using 10-fold cross-validation (10-CV). We run the 10-CV exclusively for the 21 proxy sites since the tide gauge records are short in duration and would provide relatively weak information about model performance. We present the site-specific results for only the four case study sites with the remaining sites shown in Appendix \ref{appendix_full_results}.  We evaluate the model performance based on out of sample empirical coverage and the Root Mean Squared Error (RMSE). The prediction intervals are created using posterior predictive simulations with the full error structure, i.e. $\hat{y}_{ij} \sim N(\hat{f}_{ij}, \sigma_{y_{ij}}^2 + \sigma_{{t}_{ij}}^2 + \sigma^2)$. The empirical coverage provides the percentage of occasions that the true RSL observation is within the model prediction interval (PI) for RSL. The RMSE provides insight into prediction performance in the same units as the response (meters).

The 10-CV for our full model using the 21 proxy sites obtained overall empirical coverage of 99.1\% with the 95\% prediction interval and 78.6\% with the 50\% prediction interval. These are satisfactory for a model fitted to complex data such as ours, especially with the addition of the model error. The conservative coverage values are likely a consequence of accounting for the observed measurement errors in the estimation of the prediction intervals. The RMSE for the 21 proxy sites is  0.14 m. An average out of sample error of 14 cm is reasonable given the scale and variety of the data set. 

\begin{figure}[H]
 	\centering
 	\includegraphics[width=\textwidth]{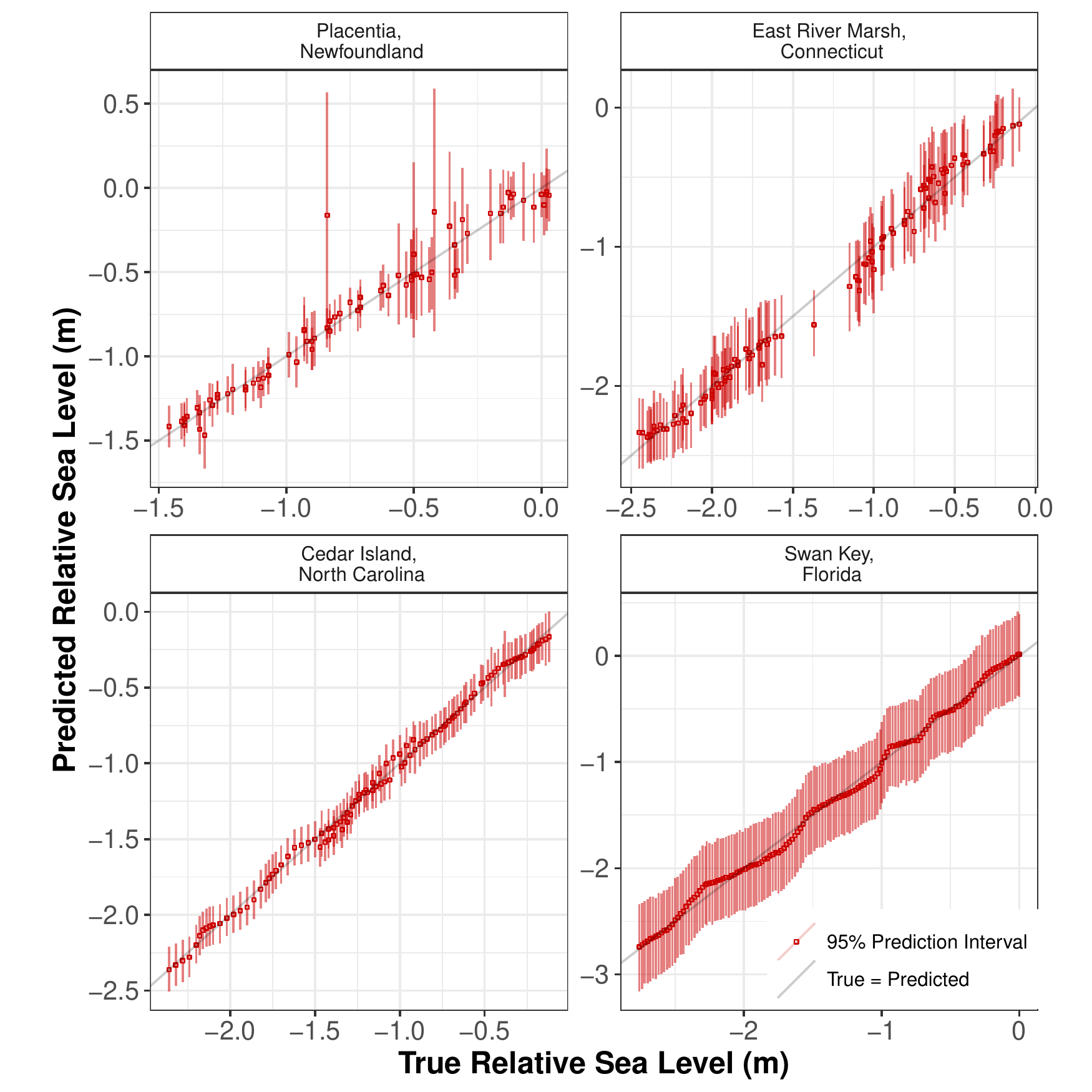}
     \caption{True vs Predicted RSL values for our case study sites at Placentia Newfoundland, East River Marsh Connecticut, Cedar Island North Carolina and Swan Key Florida using 10-fold cross validation (CV). The predicted means are the red points with a vertical 95\% prediction interval for each point. The identity line is shown in grey.}
 	\label{fig:truepred}
 \end{figure}

Figure \ref{fig:truepred} presents the true RSL observations versus the model-based RSL point estimates with 95\% prediction intervals at each site and Table \ref{Tab:model_val} provides a site-specific insight into the empirical coverage for the model and the size of the prediction intervals. Three out of the four sites have a coverage of 100\% due to the large prediction intervals arising from the bivariate uncertainties associated with the proxy data (Figure \ref{fig:SL_data}). Based on the RMSE, the best fitting case study site is Cedar Island in North Carolina, where the RMSE is 6cm (Table \ref{Tab:model_val}). At the other end of the spectrum is Swan Key Florida, where the RMSE is larger at 15 cm (Table \ref{Tab:model_val}).

\begin{table}[H]
\centering
\adjustbox{max width=\textwidth}{
\begin{tabular}{lp{21mm}p{20mm}p{21mm}p{20mm}p{10mm}}
  \hline
 Site Name & 95\% Empirical Coverage & 95\% Average PI width & 50\% Empirical Coverage & 50\% Average PI width & RMSE (m) \\ 
  \hline
    Placentia,
 Newfoundland & 0.97 & 0.36 & 0.59 & 0.12 & 0.11\\ 
  East River Marsh,
 Connecticut & 1.00 & 0.52 & 0.77 & 0.18 & 0.13\\ 
 Cedar Island,
 North Carolina & 1.00 & 0.26 & 0.78 & 0.09 & 0.06\\ 
  Swan Key,
 Florida & 1.00 & 0.77 & 1.00 & 0.26 & 0.19\\ 
   \hline
\end{tabular}}
\caption{Empirical coverage from the 10 fold cross validation and the corresponding size of the prediction intervals (PI) used for model validation for our 4 chosen sites. \label{Tab:model_val}}
\end{table}

\section{Results}\label{results}

In this section we present the results from our Bayesian hierarchical RSL model. We consider the full model fit and its decomposition into the different components of RSL, i.e., regional component, linear local component and non-linear local component.

\subsection{Full Model Fit and rate of change}

The full model fit for the four case study sites are shown in Figure \ref{fig:totalplot} (results from all 21 proxy sites are included in Appendix \ref{appendix_full_results}). The model demonstrates how the RSL varies in time at each site. Overall, the model fits the data well. The 95\% credible intervals for Swan Key Florida are larger due to the relatively large observation uncertainties at this site and the fit is notably smoother than the others. The data in Placentia, Newfoundland experiences additional variability in the observations compared with the other sites and this is reflected in a more variable total model fit.
 \begin{figure}[H]
 	\centering
 	\includegraphics[width=\textwidth]{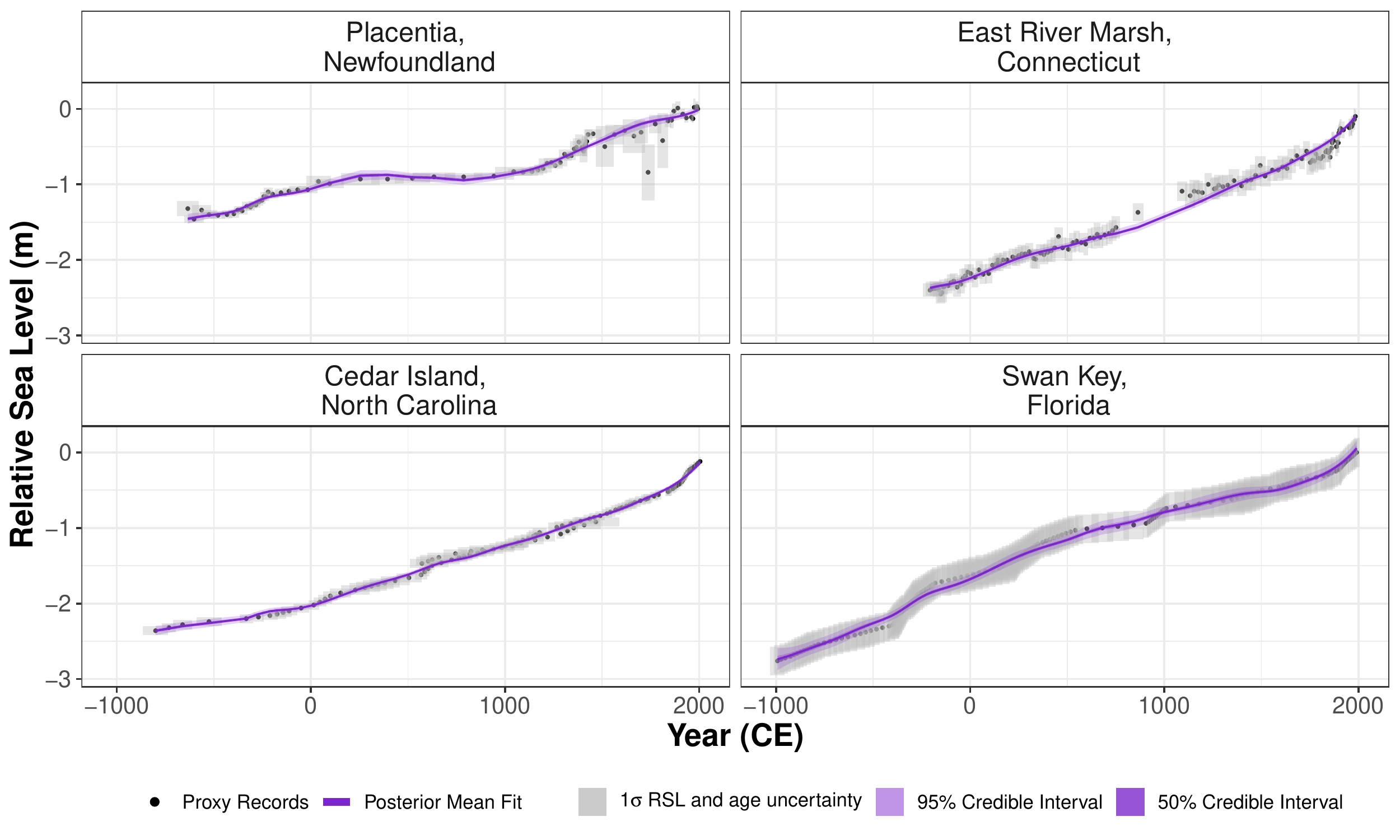}
     \caption{The noisy-input generalised additive model (NI-GAM) fit for four selected sites along the Atlantic coast of North America. The four sites include: Placentia, Newfoundland, Canada; East River Marsh, Connecticut, USA; Cedar Island, North Carolina, USA; and Swan Key, Florida, USA. The black dots and grey boxes represent the midpoint and associated uncertainty, respectively, for each proxy record. The solid purple line represents the mean of the model fit with a 95\% credible interval denoted by shading.}
 	\label{fig:totalplot}
 \end{figure}
 
\begin{figure}[H]
 	\centering
 	\includegraphics[width=\textwidth]{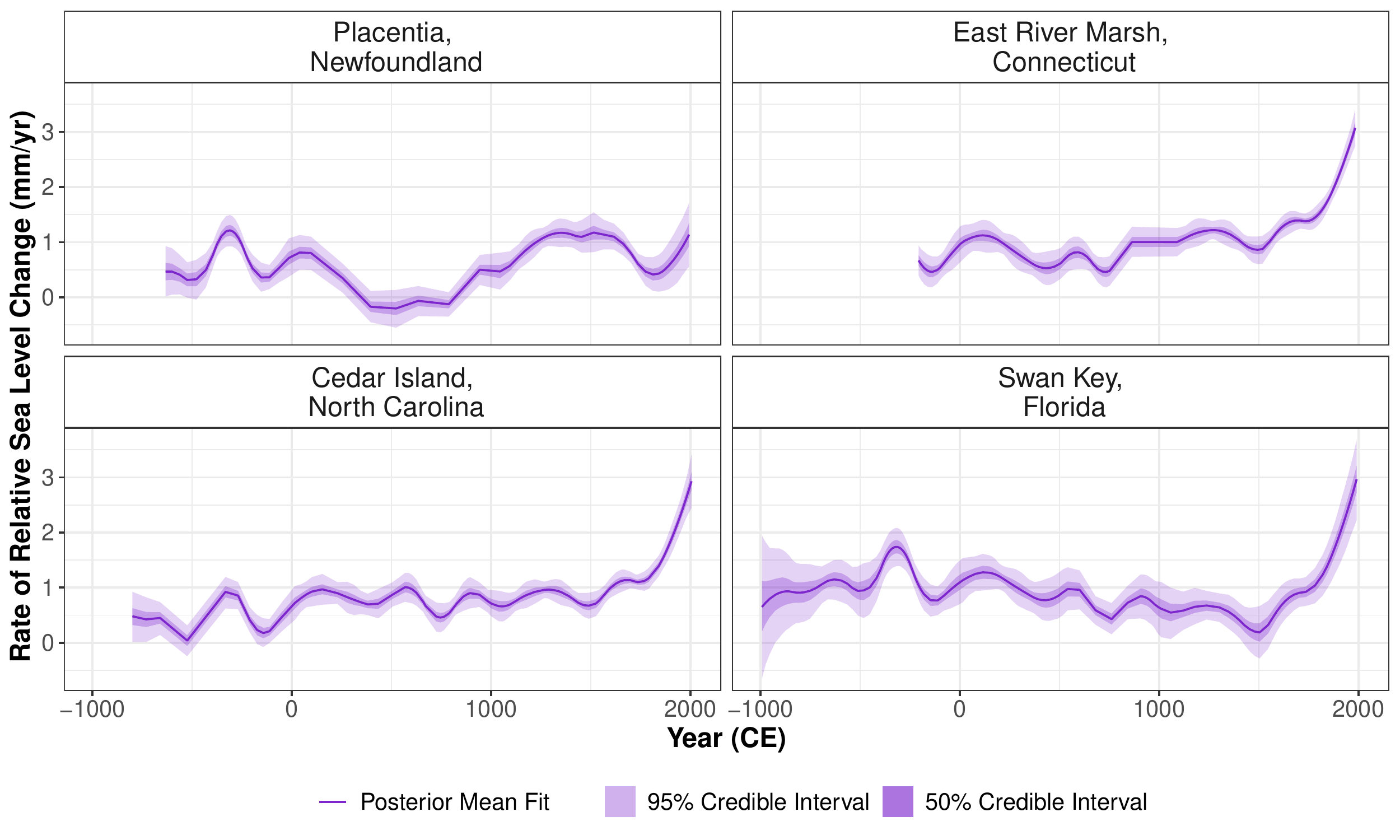}
     \caption{Rate of relative sea change found by taking the first derivative of the total model fit for four sites along the Atlantic coast of North America. The mean of the fit is the solid purple line with the dark shaded area being the 50\% credible interval and the light shaded area being the 95\% credible interval.}
 	\label{fig:siterateplot}
 \end{figure}

Figure \ref{fig:siterateplot} shows the site-specific rates of change for the case study locations calculated as described in Section \ref{subsec:model_desc}. The remaining sites are shown in Appendix \ref{appendix_full_results}. Late Holocene rates of RSL change display century to multi-century scale variability around a stable mean at each site until the 19th and 20th centuries since when the rate of rise appears unprecedented. Rates fluctuate throughout the last 2000 years but remain below 1.5mm/yr until the late 1800s in East River Marsh, Connecticut, and the early to mid 1900s in Cedar Island, North Carolina and Swan Key, Florida. The late 20th and 21st century rates at these sites are unprecedented in the last 2000 years with the most recent rates of change being 3.06 $\pm$ 0.3, 2.9 $\pm$ 0.5 and 2.9 $\pm$ 0.7 mm per year in East River Marsh Connecticut, Cedar Island North Carolina and Swan Key Florida respectively. Placentia Newfoundland does not appear to experience the same uptick in rates that the other sites do with the most recent rate being 1.21 $\pm$ 0.4 mm per year. 
 
\subsection{Examining the decomposition of RSL}
\begin{figure}[H]
 	\centering
 	\includegraphics[width=\textwidth]{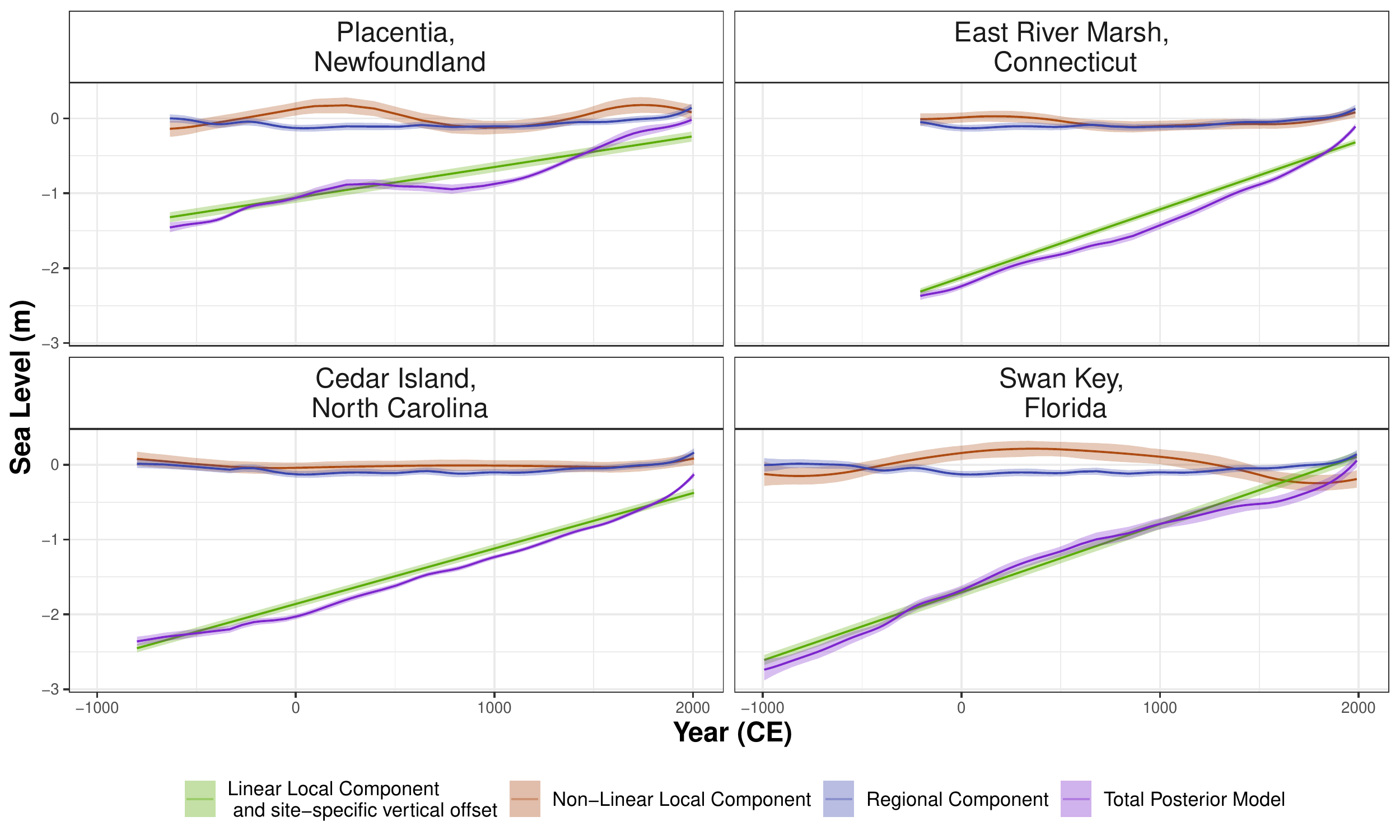}
     \caption{The decomposition of the relative sea level process level for the four sites, with solid lines indicating means and shaded areas 95\% posterior credible intervals. The blue curve represents the regional component. The brown curve represents the non-linear local component. The green line represents the site-specific vertical offset plus the linear local component. The purple line is the sum of all three components and represents the full noisy-input generalised additive model fit.}
 	\label{fig:allcomponent}
 \end{figure}

The RSL process level $f$ consists of the regional component, the linear local component, the site-specific vertical offset, and a non-linear local 
component, all as described in Section \ref{method}. Figure \ref{fig:allcomponent} illustrates the decomposition in our case study sites and provides an insight into how the components of RSL have varied over time for the Atlantic coast of North America by demonstrating the scale of the different components and how they interact over time. The total posterior model is obtained by the sum of these separate components as illustrated by the purple line in Figure \ref{fig:allcomponent}.

It is evident that the dominant driver of RSL change for these four sites until the late 1800s is the linear local component. After this interval, regional variability along the Atlantic coast of North America appears to take over and the total RSL trends at each site tend to reflect the RSL rise seen in the regional component. The non-linear local component is picking up the remaining variability and demonstrates that non-linear local effects on RSL variability are more apparent in Placentia and Swan Key compared to Cedar Island and East River Marsh.

\begin{figure}[H]
    \centering
    \begin{subfigure}{\textwidth}
        \centering
        \includegraphics[width=\textwidth]{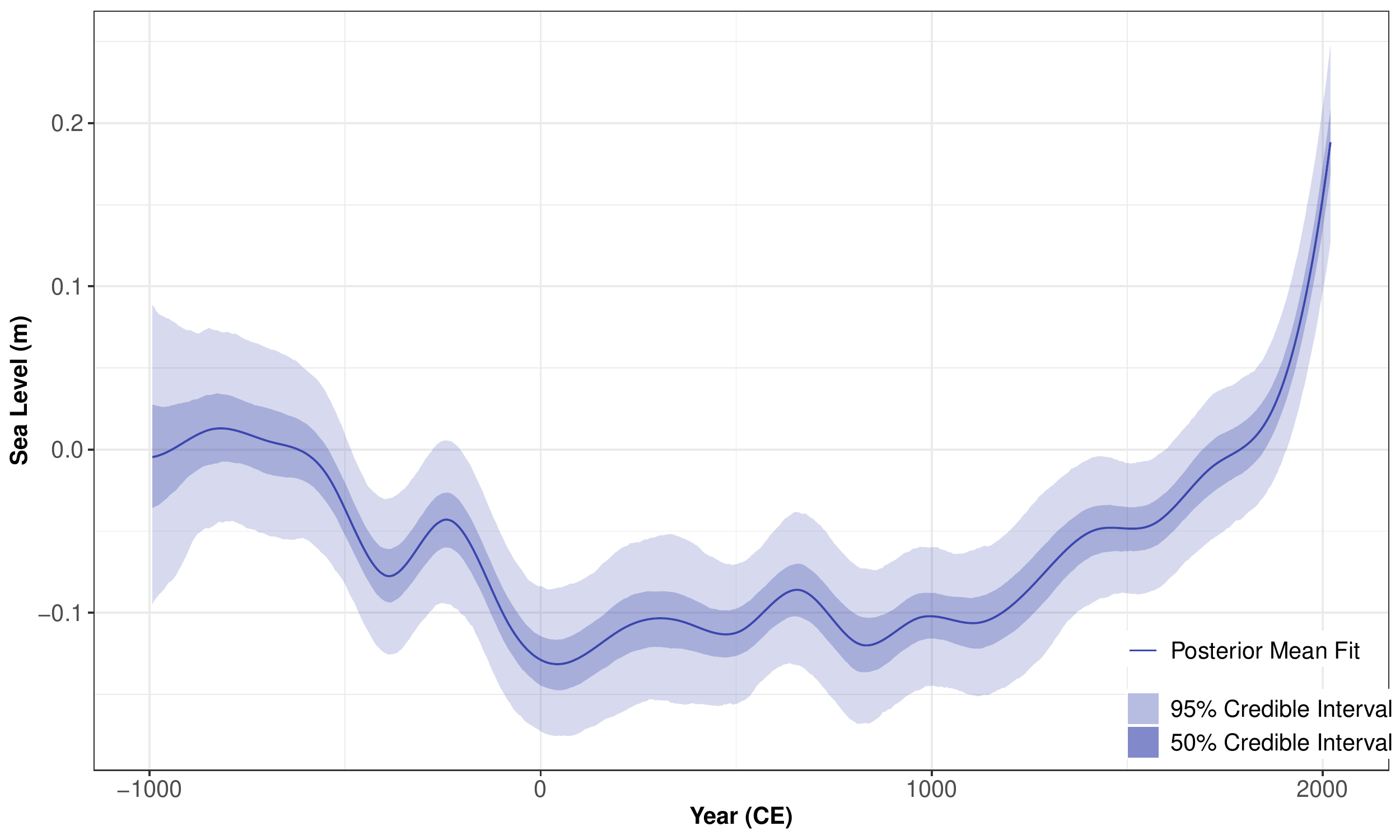}
    \end{subfigure}
    \begin{subfigure}{\textwidth}
    \includegraphics[width=\textwidth]{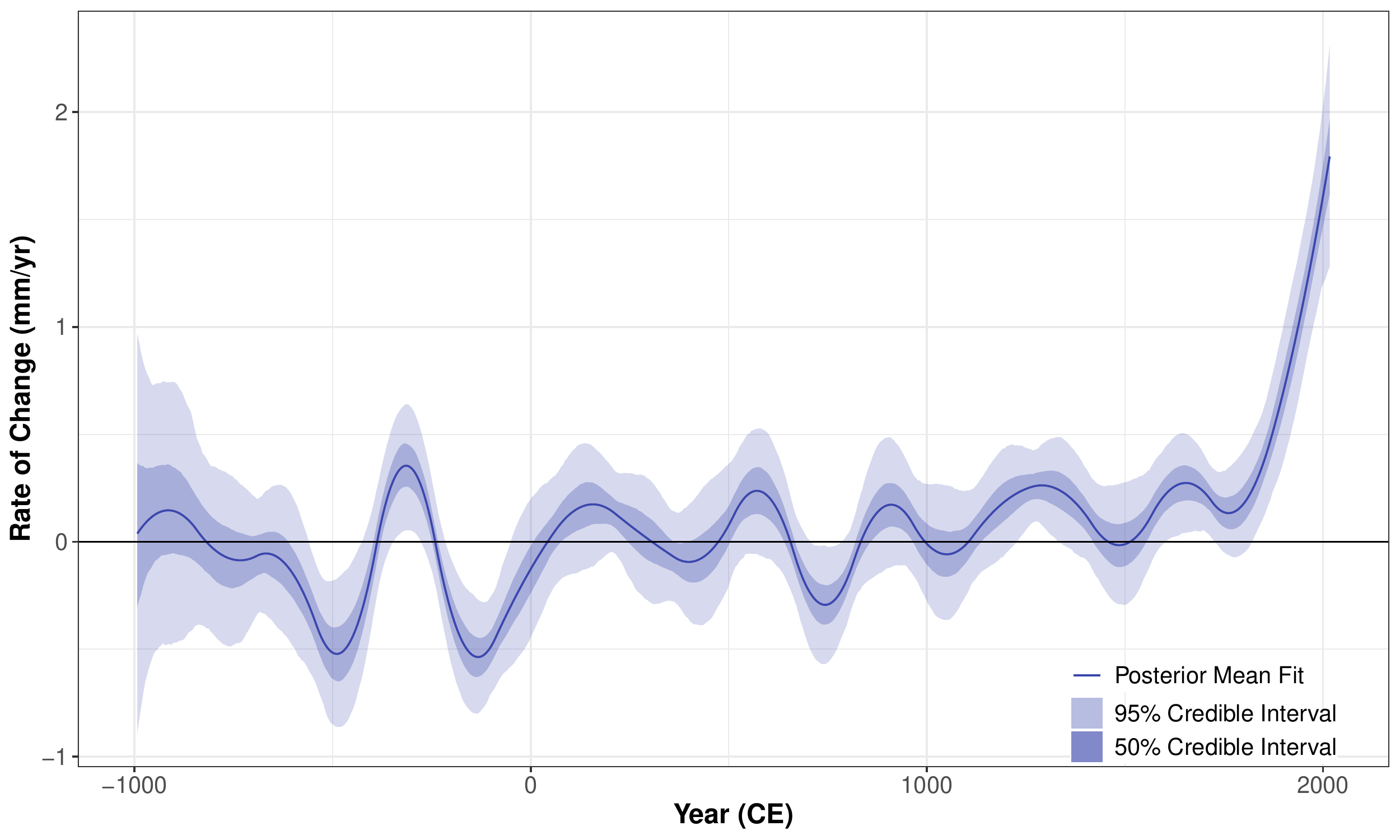}
   \end{subfigure}
    \caption{The noisy-input generalised additive model (NI-GAM) results for (a) the regional component and (b) the regional rate of change component. (a) The regional component mean model fit  represented with a solid line and the shading indicating the 50\% credible interval in dark blue and 95\% credible interval in light blue. The $y$-axis is the sea level in m with the $x$-axis representing the time across the last 3000 years for the Atlantic coast of North America.
     (b) Rate of Change for the regional component for the Atlantic coast of North America with the solid line representing the mean of the fit, the dark blue shaded area representing the 50\% credible interval and the light blue shaded area representing the 95\% credible interval. The $y$-axis is the instantaneous rate of change of regional sea level in mm per year.}
 	\label{fig:regionalplots}
 \end{figure}

Figure \ref{fig:regionalplots}(a) shows the regional component  (common to all sites) for the 21 proxy sites and 66 tide gauge sites along the Atlantic coast of North America. As a reminder, the regional component is represented with a spline in time which is common across all sites. Prior to 0 CE, sea level fluctuated from 0.01 m to -0.13 m. From 0 CE to 1200 CE, variability of sea level reduced ranging from -0.12 m to -0.09 m. Following 1200 CE, a sharp increase in sea level can be seen with brief periods of stability from ~ 1410 CE to ~1560 CE and from ~1800 CE to ~1840 CE. After 1800 CE, sea levels are consistently rising and the most dramatic increase can been seen from the mid-1800s until the present day. Figure \ref{fig:regionalpostsamples} in Appendix \ref{appendix_full_results} demonstrates the underlying behaviour of the posterior samples for the regional component. 
Figure \ref{fig:regionalplots}(b) shows the rate of change for the regional component along the Atlantic coast of North America. Rates fluctuate around 0 CE between -990 CE and ~1800 CE after which a continuous increase can be see from ~1800 CE onwards. The rate from the late 20th century is unprecedented when compared with the last 3000 years and is estimated to be 1.8 $\pm$ 0.5 mm per year. This 20th regional sea-level rate depicts patterns over multi-decadal to centennial timescales due to the resolution limits and natural time-averaging of proxy reconstructions and decadally averaged of tide-gauge data.

The linear local component is represented with a random slope effect as described in Section \ref{method}. As stated in Section \ref{drivers_RSL}, this parameter removes a long term variation driven principally (but perhaps not exclusively) by GIA. Table \ref{Tab:GIARates} compares our empirically estimated values of this parameter for the proxy record sites in mm per year prior to 1800 CE. It is clear that there is wide variability in these values between sites with areas such as Swan Key Florida and East River Marsh Connecticut experiencing rates of 0.91 mm/yr whereas Placentia Newfoundland has just 0.41 mm/yr. To show a comparison with physical model-based GIA rates, Table \ref{Tab:GIARates} presents values obtained from the ICE5G - VM2-90 Earth-ice model \citep{Peltier2004}. It is evident that the data-driven rates and the GIA-model rates differ, with Swan Key Florida experiencing the greatest difference of 0.8mm/yr. Whereas, East River Marsh Connecticut has similar rates with a difference of only 0.05mm/yr.

\begin{table}[H]
    \centering
    \begin{tabular}{p{0.4\linewidth}p{0.2\linewidth}p{0.2\linewidth}}
    \hline
     & Empirical Rate prior to 1800 CE [mm/yr] & ICE5G-VM2-90 Earth-ice GIA rate [mm/yr] \citep{Peltier2004} \\
     \hline
        Placentia, Newfoundland & 0.41 & 0.21
        \\
        East River Marsh, Connecticut & 0.91 & 0.96
        \\
        Cedar Island, North Carolina & 0.74 & 0.69  \\
        Swan Key, Florida & 0.91 & 0.11 \\
    \hline
    \end{tabular}
    \caption{Linear local component for our four sites along the Atlantic coast of North America given in mm per year. The empirical rates represent the rate estimated from the data prior to 1800 CE, which is used to inform the priors for the linear local component \citep{2K_2019}. ICE5G-VM2-90 Earth-ice GIA rate is from an Earth-ice physical model \citep{Peltier2004}. \label{Tab:GIARates}}
\end{table}

\begin{figure}[H]
 	\centering
 	\includegraphics[width=\textwidth]{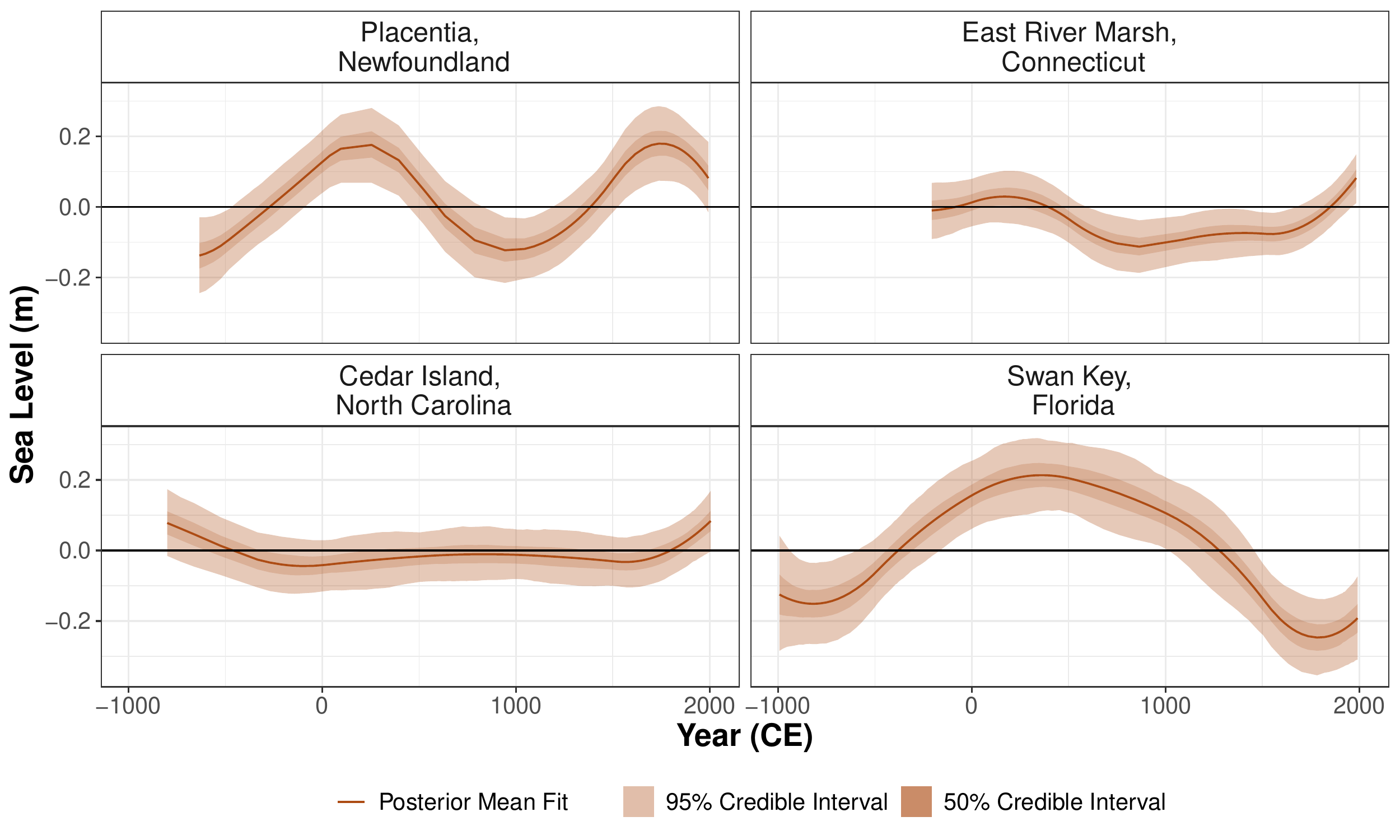}
     \caption{The non-linear local component for our four sites along the Atlantic coast of North America. The y-axis represents sea level in meters. The brown solid line represents the mean of the model fit with the 50 \% credible interval in dark brown shading and 95\% credible interval in the light brown shading.}
 	\label{fig:localplot}
 \end{figure}

Figure \ref{fig:localplot} shows our non-linear local component that represents the spatially structured behaviour specific to each site. There are clearly different patterns of non-linear local sea-level change, which is to be expected given that the common source of variation across all sites has been captured by the regional component. Placentia Newfoundland and Swan Key Florida show non-linear local variations in sea level ranging from 0.19 m to -0.12 m and 0.21 m to -0.25 m respectively.  On the other hand, Cedar Island North Carolina and East River Marsh Connecticut do not experience this level of variability with sea levels in the non-linear local component fluctuating close to zero.

\section{Discussion}\label{discussion}
RSL change is the net result of multiple physical processes within the oceans, atmosphere, and solid Earth, that can alter the height of the land and/or sea surface \citep{Church2001}. The importance of specific processes varies markedly across space and through time, giving rise to a complex and evolving pattern of RSL change. Tide-gauge data and proxy records contain information about these processes and the sea-level community requires advanced statistical tools to decompose the net RSL signal into contributions from physical processes while accounting for uncertainties in the underlying data. The need for RSL decomposition motivated current modeling strategies such as the K16 model and our approach presented here.  

Our approach provides a more computationally efficient method for decomposing the RSL signal. The process level of our model utilises a spatio-temporal field decomposed into: a regional component; a linear local component; and a non-linear local component. In contrast to K16 which uses GPs, we use splines to examine the different drivers of RSL. This is due to the computational complexity associated with the likelihood computation for a Gaussian Processes being of $O(n^3)$ where $n$ is the number of data points. In contrast, the likelihood computation for the equivalent spline with pre-computed basis functions is just $O(n)$ \citep{wood_2017}. Thus our model can be fitted quicker than K16, allowing for further checks on the performance of our model. The model validations presented in Section \ref{modelval} highlight this and we are confident that the NI-GAM is effectively capturing the different components of RSL along the Atlantic coast of North America. In addition, the construction of our GAM using spline basis functions and random effects allows for easy interpretability without the need for covariance matrices and correlation functions \citep{Porcu2021}. We can efficiently model late Holocene RSL changes along the Atlantic coast of North America and the interpretability of GAMs allows for these changes to be easily examined (Figure \ref{fig:totalplot}).

Our approach attempts to deviate from the Empirical Bayesian framework as implemented by K16 and related models. \citet{Piecuch2017} demonstrated that Empirical Bayesian methodologies can underestimate uncertainty when examining historic sea-level change along the Atlantic coast of North America. However, we recognise the difficulty of a fully Bayesian approach due to the confounding nature of the regional, linear local, and non-linear local components. Instead we opted to take advantage of the two-step fitting required by our use of the noisy-input method to take account of age errors. The first step of the modeling procedure obtains posterior distributions for the regional component and the site-specific vertical offset. The second step uses the resulting posterior estimates and uncertainties to inform the priors for the remaining linear local and non-linear local components, and the extra measurement variance contribution from the age uncertainties. The first step can be thought of as estimating the main component of our model: the regional RSL curve, with the second step designed to decompose the residuals and ensure the uncertainty is properly calibrated. Our modelling strategy avoids fixing process model parameters, and severe confounding that would occur were we to fit the model in one step.

Considering the individual RSL components, pre-anthropogenic \citep[before 1800 CE;][]{2K_2019} RSL change along the Atlantic coast of North America is dominated by the linear local component which is principally capturing the contribution from ongoing GIA. However, there are some notable differences between the empirically-estimated rates obtained from our models and the GIA rates obtained from the ICE5G-VM2-90 Earth-ice physical model (Table \ref{Tab:GIARates}). There are several possible explanations for these discrepancies. First, a single Earth-ice model generates GIA predictions from a specific representation of the solid Earth (e.g., mantle viscosity and lithospheric thickness parameters) and history of deglaciation. It is unlikely that any single Earth-ice model will perfectly estimate GIA at all places and all times because the parameters are uncertain and may vary by location \citep{Roy2015}. In particular, locations close to the margins of former ice sheets (such as Newfoundland) may exhibit particularly pronounced differences in GIA estimated by different Earth-ice models. Systematic difference between RSL predicted by specific Earth-ice models and proxy reconstructions on Holocene timescales is well documented in eastern North America \citep{vacchi2018} and elsewhere \citep{shennan2018}. One such example is Placentia Newfoundland where our empirically-estimated rate and the Earth-ice physical model GIA rate differ by 0.1 mm/yr (Table \ref{Tab:GIARates}). Second, physical Earth-ice models only estimate the contribution from GIA, while the empirical approach captures contributions from other processes such as vertical land motion from tectonic process which may also be a linear driver of RSL change on the timescales under consideration. Although these non-GIA processes may be small on the passive margin of the Atlantic coast of North America, they are also unlikely to be zero at all sites. For example, \citet{Khan2022} identified an anomalously high rate of rise at Swan Key Florida compared to nearby Snipe Key Florida and proposed that dissolution of the carbonate bedrock beneath the site resulted in an additional approximately linear component of RSL rise. This is highlighted in Table \ref{Tab:GIARates} where Swan Key Florida exhibits a large difference between the empirical rate of 0.91 mm/yr compared with the ICE5G-VM2-90 GIA rate of 0.11 mm/yr.

After $\sim1900$ CE the regional component dominates and we see the regional rate of change increase markedly from 0.7 $\pm$ 0.5 mm/yr in 1902 to 1.8 $\pm$ 0.5 mm/yr at the end of the 20$^{th}$ century (Figure \ref{fig:regionalplots}(b)). This change is the result of anthropogenic forcing of the climate system \citep{2K_2019}, which drove sea-level rise through thermosteric and barystatic processes \citep{frederikse2020}. Our estimate of regional sea-level rise during the 20$^{th}$ century represents trends sustained on multi-decadal to centennial timescales because of the natural time-averaging and resolution limits of the proxy reconstructions and our decadal average of tide gauge measurements. Despite our analysis being limited to the Atlantic coast of North America, our estimated rate is comparable to century-scale estimates generated using only tide gauge data \citep{Hay2015,frederikse2020} and the K16 statistical model \citep{Kopp2016,walker2022timing}.

The diverse trends captured by the non-linear local component, as shown in Figure \ref{fig:localplot}, highlight the important influence site-specific processes can have on the RSL. At Placentia Newfoundland the non-linear local component experiences large fluctuations with maximum peaks reaching values of $\sim$ 0.19m at around 250 CE and 1775 CE and minimum troughs of -0.12 m at around -450 CE and 1050 CE. This is a particularly pronounced degree of variability. The original study of the site by \citet{Kemp2018} recognised that the geomorphology at Placentia rendered it sensitive to site-specific RSL change due to the position of the salt marsh. The salt marsh is separated from the open ocean by a narrow inlet which is likely prone to opening and closing of the dynamic sediment barrier. In contrast, the East River Marsh record was generated exclusively through sediment in direct contact with bedrock to negate the potential influence of sediment compression as a driver of RSL change \citep{Kemp2015}. This contrast is reflected in our estimate of the non-linear local component where variability is present with a slight increase in sea level followed by a fall at around 100 CE and from 650 CE onwards an increase. Thus the component is non-zero due to the presence of other processes that can affect individual sites or groups of sites \citep[e.g., dynamic sea level change][]{Kemp2015}.

There are a number of potential extensions to the NI-GAM model which have not been addressed in our paper. A future aim is to extend NI-GAM further to larger regions, e.g. North Atlantic, or potentially to examine global RSL trends. This poses a challenge as the network of proxy records and tide gauges is non-uniformly spread and biased to coastal regions in the Northern hemisphere. Previous attempts to resolve this spatial bias have used a variety of techniques, \citep[e.g.][]{jevrejeva2008,Wenzel2010,Church2011,Hay2015,Dangendorf2017,Berrett2020}, yet have mostly focused on instrumental data from tide gauges and satellites. Models like K16 and its extensions \citep[e.g.][]{Khan2017,Kemp2018,Walker2021} created a global component which may give insight into the changes in sea level common across many sites. Therefore, our model would require more components and further solutions to additional potential confounding issues. Yet, NI-GAM is an extendable modeling approach due the flexible structures of spline-GAMs and the Bayesian framework which allows for the inclusion of informed priors from future RSL analyses. Our modelling strategy is of course not limited to RSL changes. Rather it has the potential to be expanded to other areas of research that require the decomposition of a signal into different components that vary in time and space with complex measurement errors. One such example would be investigating historic temperature trends at a local and regional level to gauge the components that alter temperature spatially and temporally.

\appendix 

\section{Appendix}\label{appendix_full_data}

\subsection*{Data}\label{full_data}
\begin{sidewaysfigure}
\begin{figure}[H]
 	\centering
 	\includegraphics[width=\textwidth]{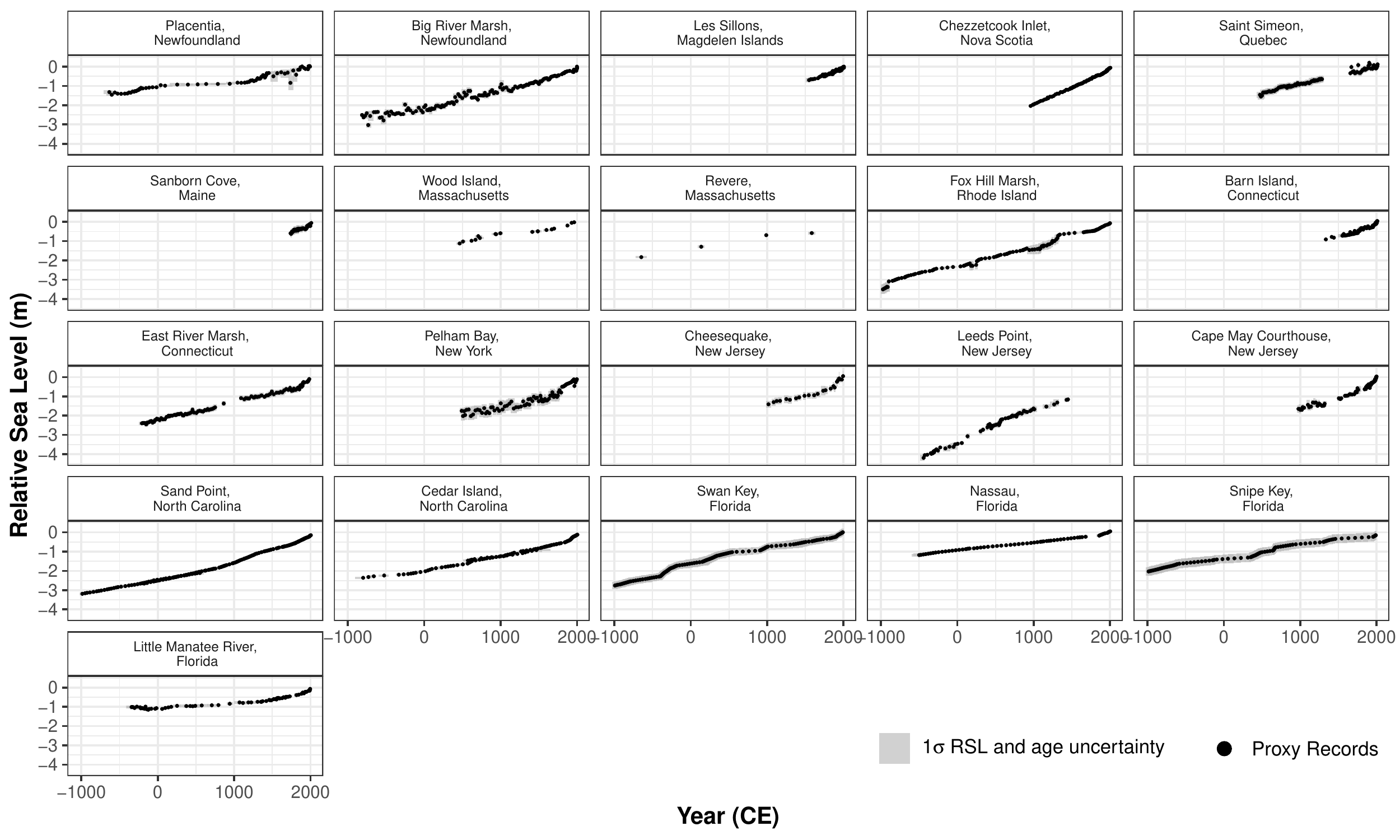}
     \caption{Proxy records for 21 sites along Atlantic coast of North America. The grey boxes represents the 1 $\sigma$ uncertainty in RSL and age. The black points represents the midpoint of the uncertainty boxes which we use as the input of our data.}
     \label{fig:SLdata_full}
 \end{figure}
\end{sidewaysfigure}

 Table \ref{Tab:full_data_reference} provides a list of all the proxy record sites used along the Atlantic coast of North America in our model and Figure \ref{fig:SLdata_full} represents the proxy data associated with the 21 proxy record sites used in our model with the grey boxes the 1 $\sigma$ uncertainty in the age and RSL value and the black dots the midpoint of the uncertainty box. Table \ref{Tab:full_data_reference} gives the references associated with each proxy data site and more information in regard to data collection can be sourced here. In addition, Table \ref{Tab:full_data_reference} contains the GIA rate used to inform the prior for the linear local component calculated using a linear regression for the data prior to 1800 CE for each site \citep{2K_2019}. This provides an estimate for the rebounding effect of the tectonic plate after a glacier melts \citep{Whitehouse2018}. In previous models, physical GIA models are used to inform the prior for the linear local component. \citet{Peltier2004} developed the ICE5G - VM2-90 Earth - ice which provides a GIA rate for each site. We carried out comparison between these techniques however, the data driven GIA rates were our preferred choice for the proxy records.

Table \ref{Tab:full_TG_data_reference} gives the reference name associated with each tide-gauge data site and its' location along North America's Atlantic coast using PSMSL database \citep{Holgate_PSMSL2013}. For each tide gauge site, the associated GIA rate is provided which is obtained using the ICE5G - VM2-90 Earth - ice developed by \citet{Peltier2004}. The uncertainty associated with these values is selected to be 0.3 mm per year based on study carried out by \citet{Engelhart2009}.

\begin{table}[H]
\centering
\adjustbox{max width=\textwidth}{
\begin{tabular}{p{50mm}lp{20mm}p{20mm}}
  \hline 
  Reference & Site Name & Empirical rate prior to 1800 CE (mm/yr) & ICE5G GIA rate (mm/yr) \citep{Peltier2004} \\ 
  \hline  
 \cite{Donnelly2004,Gehrels2020} & Barn Island,
 Connecticut & 1.00 & 0.99\\ 
  \cite{Kemp2015, Stearns2017} & East River Marsh,
 Connecticut & 0.91 & 0.96 \\ 
  \cite{Kemp2014} & Nassau,
 Florida & 0.42 & 0.28\\ 
  \cite{Donnelly2006} & Revere,
 Massachusetts & 0.58 & 0.40 \\ 
\cite{Kemp2011} & Wood Island,
 Massachusetts & 0.52 & 0.40\\ 
  \cite{Kemp2011, Kemp2017} & Sand Point,
 North Carolina & 0.97 & 0.69\\ 
  \cite{Kemp2011, Kemp2017} & Cedar Island,
 North Carolina & 0.74 & 0.69\\  \cite{Kemp2013Sea-levelUSA,Cahill2016} & Cape May Courthouse,
 New Jersey & 1.19 & 1.24\\  \cite{Kemp2013Sea-levelUSA, Cahill2016} & Leeds Point,
 New Jersey & 1.69 & 1.41\\ 
   \cite{Barnett2017}  & Les Sillons,
 Magdelen Islands & 1.23 & 2.20\\ 
   \cite{Gerlach2017}  & Little Manatee River,
 Florida & 0.28 & 0.14\\ 
    \cite{Kemp2018} & Big River Marsh,
 Newfoundland & 0.87 & 0.60\\ 
   \cite{Kemp2018} & Placentia,
 Newfoundland & 0.41 & 0.21\\ 
   \cite{Barnett2019} & Saint Simeon,
 Quebec & 0.93 & 2.33\\ 
   \cite{Gehrels2020} & Chezzetcook Inlet,
 Nova Scotia & 1.76 & 0.63\\ 
   \cite{Gehrels2020} & Sanborn Cove,
 Maine & 3.23 & 0.08\\ 
  \cite{Khan2022} & Snipe Key,
 Florida & 0.66 & 0.13\\ 
  \cite{Khan2022} & Swan Key,
 Florida & 0.90 & 0.11\\ 
  \cite{Walker2021} & Cheesequake,
 New Jersey & 0.85 & 1.31\\ 
   \cite{Kemp2017,Stearns2017} & Pelham Bay,
 New York & 0.8 & 1.31\\ 
  \cite{Stearns2017} & Fox Hill Marsh, Rhode Island & 1.00 & 1.07\\
   \hline
\end{tabular}}
\caption{The 21 data sites used in our model and the associated reference for each location. The Site Name are a combination of the site-specific name and the corresponding state. In addition, a comparison is made between the GIA rates we use from the data and GIA rates from physical models such as the ICE5G - VM2-90 Earth-ice model by \cite{Peltier2004}. \label{Tab:full_data_reference}}
\end{table}

\begin{table}[H]
\centering
\tiny
\adjustbox{width=\textwidth}{
  \begin{tabular}{cclp{20mm}}
    \hline
Longitude & Latitude & Site Name &  ICE5G-VM2-90 GIA rate (mm/yr) \citep{Peltier2004} \\ 
  \hline
-54.00 & 47.30 & ARGENTIA & 0.21 \\ 
  -74.40 & 39.40 & ATLANTICCITY & 1.41 \\ 
  -68.20 & 44.40 & BARHARBOR,FRENCHMANBAY,ME & -0.11 \\ 
  -76.70 & 34.70 & BEAUFORT,NORTHCAROLINA & 0.61 \\ 
  -63.60 & 44.70 & BEDFORDINSTITUTE & 0.49 \\ 
  -65.80 & 47.90 & BELLEDUNE & 1.98 \\ 
  -74.10 & 40.60 & BERGENPOINT,STATENIS. & 1.31 \\ 
  -53.10 & 48.70 & BONAVISTA & 0.36 \\ 
  -71.10 & 42.40 & BOSTON & 0.40 \\ 
  -64.00 & 44.70 & BOUTILIERPOINT & 0.49 \\ 
  -73.20 & 41.20 & BRIDGEPORT & 0.96 \\ 
  -70.60 & 41.70 & BUZZARDSBAY & 1.07 \\ 
  -61.90 & 47.40 & CAPAUXMEULES & 2.20 \\ 
  -75.60 & 35.20 & CAPEHATTERAS,NORTHCAROLINA & 0.69 \\ 
  -75.00 & 39.00 & CAPEMAY & 1.24 \\ 
  -83.00 & 29.10 & CEDARKEYI & 0.24 \\ 
  -63.10 & 46.20 & CHARLOTTETOWN & 1.67 \\ 
  -82.80 & 28.00 & CLEARWATERBEACH & 0.25 \\ 
  -67.20 & 44.60 & CUTLER & 0.08 \\ 
  -67.30 & 44.60 & CUTLERII & 0.08 \\ 
  -66.40 & 48.10 & DALHOUSIE & 2.12 \\ 
  -75.70 & 36.20 & DUCKPIEROUTSIDE & 0.67 \\ 
  -67.00 & 44.90 & EASTPORT & 0.08 \\ 
  -81.50 & 30.70 & FERNANDINABEACH & 0.42 \\ 
  -81.90 & 26.60 & FORTMYERS & 0.13 \\ 
  -70.70 & 43.10 & FORTPOINT,NEWCASTLEISLAND & -0.30 \\ 
  -63.60 & 44.70 & HALIFAX & 0.49 \\ 
  -80.10 & 25.90 & HAULOVERPIER & 0.11 \\ 
  -75.10 & 38.60 & INDIANRIVERINLET & 1.13 \\ 
  -81.60 & 30.40 & JACKSONVILLE & 0.28 \\ 
  -81.00 & 24.70 & KEYCOLONYBEACH & 0.07 \\ 
  -81.80 & 24.60 & KEYWEST & 0.13 \\ 
  -73.80 & 40.80 & KINGSPOINT,NEWYORK & 1.31 \\ 
  -58.40 & 49.10 & LARKHARBOUR & -0.82 \\ 
  -75.10 & 38.80 & LEWES(BREAKWATERHARBOR) & 1.13 \\ 
  -64.90 & 47.10 & LOWERESCUMINAC & 2.24 \\ 
  -81.00 & 24.70 & MARATHONSHORES & 0.07 \\ 
  -81.40 & 30.40 & MAYPORT & 0.42 \\ 
  -81.40 & 30.40 & MAYPORT(BARPILOTSDOCK),FLORIDA & 0.42 \\ 
  -80.10 & 25.80 & MIAMIBEACH & 0.11 \\ 
  -72.00 & 41.00 & MONTAUK & 0.99 \\ 
  -76.70 & 34.70 & MOREHEADCITY & 0.61 \\ 
  -72.10 & 41.40 & NEWLONDON & 0.99 \\ 
  -71.30 & 41.50 & NEWPORT & 1.07 \\ 
  -73.80 & 40.90 & NEWROCHELLE & 1.31 \\ 
  -74.00 & 40.70 & NEWYORK(THEBATTERY) & 1.31 \\ 
  -60.20 & 46.20 & NORTHSYDNEY & 1.51 \\ 
  -75.50 & 35.80 & OREGONINLETMARINA,NORTHCAROLINA & 0.91 \\ 
  -75.10 & 39.90 & PHILADELPHIA(PIER9N) & 1.24 \\ 
  -62.70 & 45.70 & PICTOU & 0.93 \\ 
  -72.20 & 41.20 & PLUMISLAND & 0.99 \\ 
  -59.10 & 47.60 & PORTAUXBASQUES & 1.19 \\ 
  -73.10 & 41.00 & PORTJEFFERSON & 0.96 \\ 
  -76.30 & 36.80 & PORTSMOUTH(NORFOLKNAVYYARD) & 0.67 \\ 
  -71.40 & 41.80 & PROVIDENCE(STATEPIER) & 1.07 \\ 
  -64.40 & 49.00 & RIVIERE-AU-RENARD & 1.12 \\ 
  -63.30 & 46.50 & RUSTICO & 1.67 \\ 
  -70.50 & 41.80 & SANDWICHMARINA,CAPECODCANALENTRANCE & 1.20 \\ 
  -74.00 & 40.50 & SANDYHOOK & 1.31 \\ 
  -52.70 & 47.60 & ST.JOHN'S,NFLD. & 0.41 \\ 
  -82.60 & 27.80 & ST.PETERSBURG & 0.26 \\ 
  -55.40 & 46.90 & STLAWRENCE & 0.59 \\ 
  -81.10 & 24.70 & VACAKEY & 0.07 \\ 
  -80.20 & 25.70 & VIRGINIAKEY,FL & 0.11 \\ 
  -73.80 & 40.80 & WILLETSPOINT & 1.31 \\ 
  -70.70 & 41.50 & WOODSHOLE(OCEAN.INST.) & 1.07 \\ 
   \hline
\end{tabular}}
\caption{The 66 tide-gauge data sites and their geographical coordinates used in our model from \citet[PSMSL][]{Holgate_PSMSL2013}. Also, the GIA rate for each tide gauge site is provided using the ICE5G - VM2-90 Earth - ice physical model developed by \citet{Peltier2004}. \label{Tab:full_TG_data_reference}}
\end{table}

\section{Appendix}\label{appendix_full_results}

\subsection*{Model results and Validations for the full data set}
In this section, we will present the results from our full dataset of 21 proxy sites and 66 tide gauges. In addition, we present the results from the 10-fold cross validation using the 21 proxy sites. 

\subsubsection*{Results for full dataset}
\begin{figure}[H]
 	\centering
 	\includegraphics[width=\textwidth]{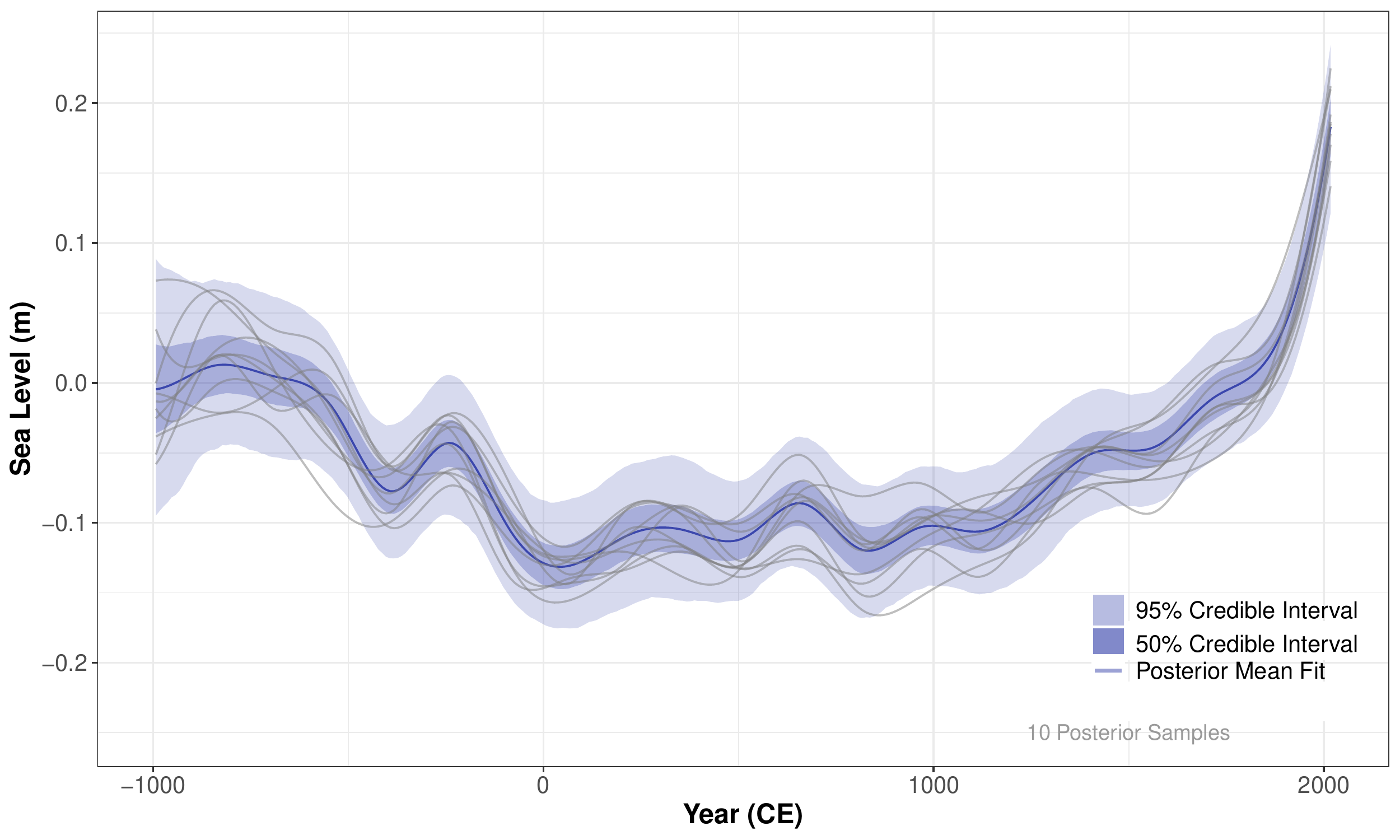}
     \caption{Regional component of the noisy-input generalised additive model using 21 proxy sites and 66 tide gauge sites along the Atlantic coast of North America. The dark blue line highlights the mean posterior model fit and the dark blue shading indicated the 50\% credible interval and the lighter blue shading is the 95\% credible interval. The grey lines represent 10 posterior samples to demonstrate that the samples are parallel.}
 	\label{fig:regionalpostsamples}
 \end{figure}

\begin{sidewaysfigure}
 \begin{figure}[H]
 	\centering
 	\includegraphics[width=\textwidth]{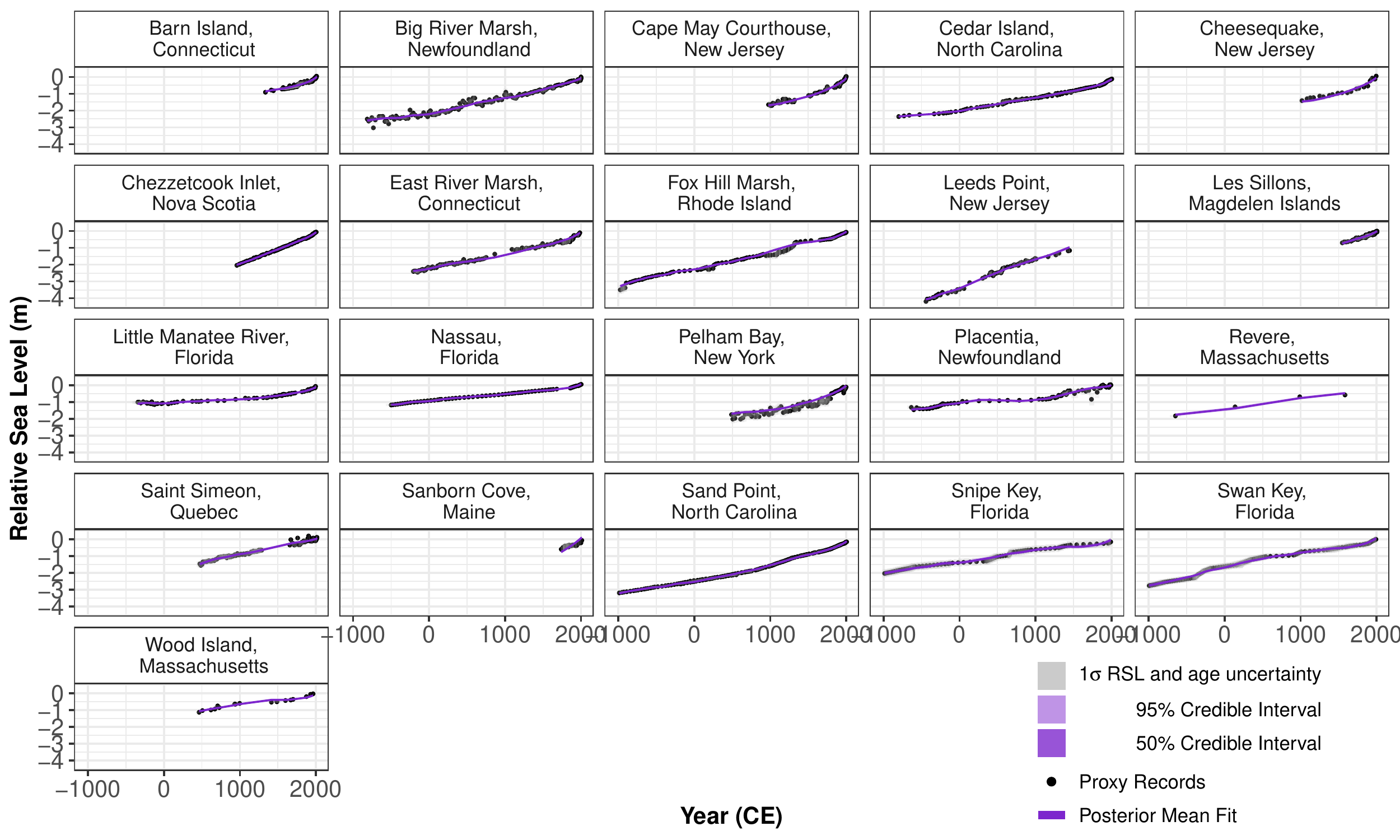}
     \caption{The noisy-input generalised additive model (NI-GAM) fit for 21 proxy sites along the Atlantic coast of North America.  The black dots and grey boxes represent the midpoint and associated uncertainty, respectively, for each proxy record. The solid purple line represents the mean of the model fit with a 95\% credible interval denoted by shading.}
 	\label{fig:totalfulldataproxy}
 \end{figure}
\end{sidewaysfigure}

\begin{sidewaysfigure}
 \begin{figure}[H]
 	\centering
 	\includegraphics[width=\textwidth]{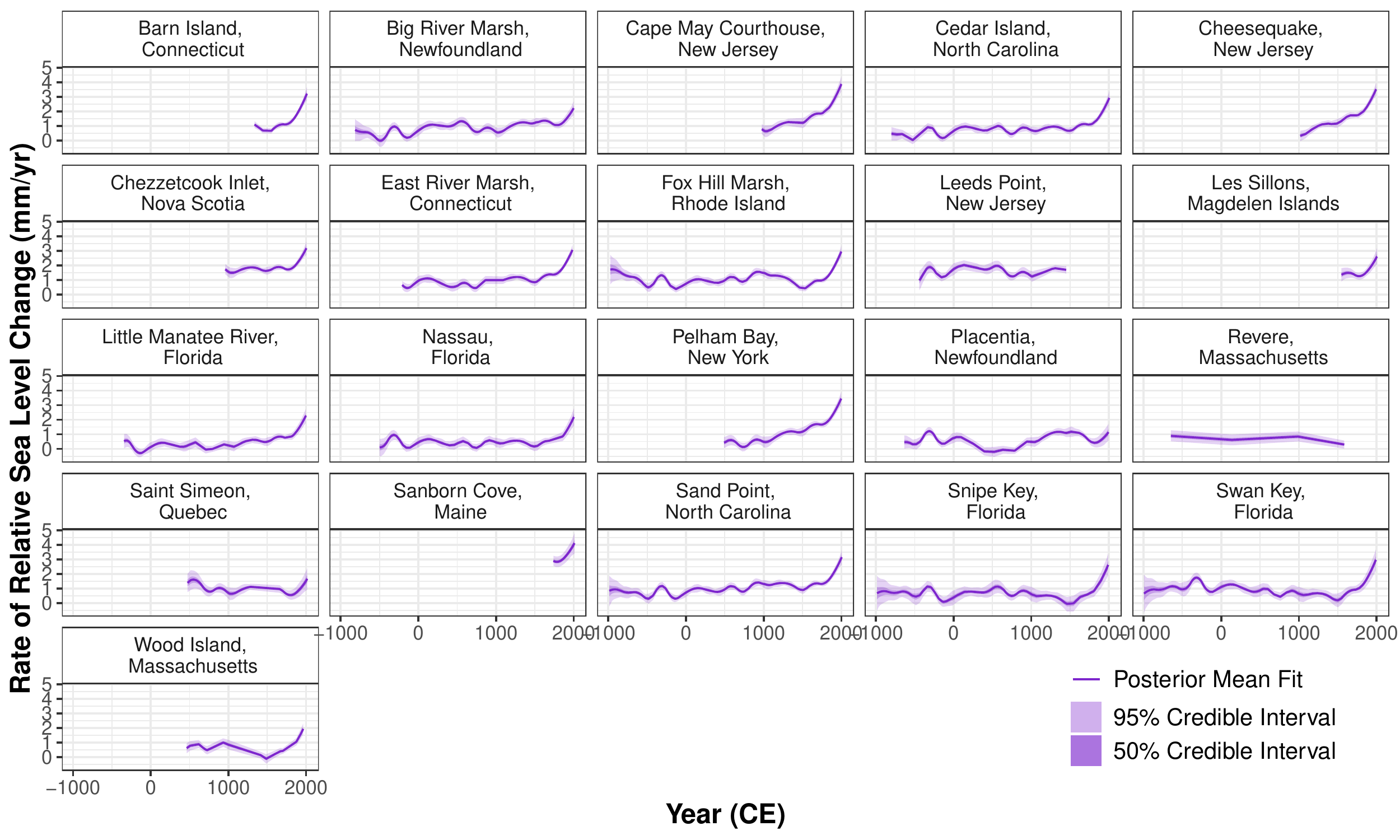}
     \caption{Rate of relative sea change found by taking the first derivative of the total model fit for 21 proxy sites along the Atlantic coast of North America. The mean of the fit is the solid purple line with the dark shaded area being the 50\% credible interval and the light shaded area being the 95\% credible interval.}
 	\label{fig:ratefulldata}
 \end{figure}
\end{sidewaysfigure}

\begin{sidewaysfigure}
 \begin{figure}[H]
 	\centering
 	\includegraphics[width=\textwidth]{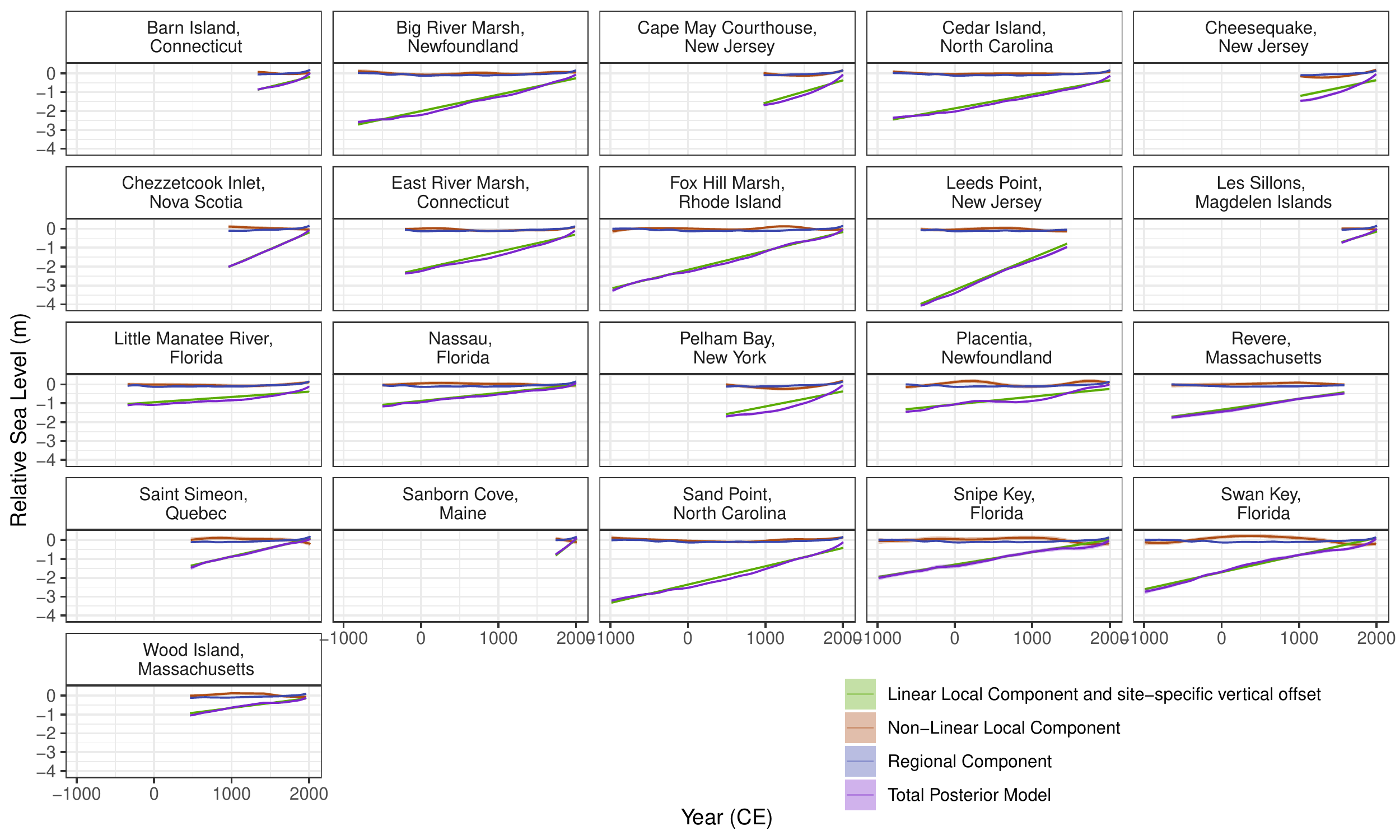}
     \caption{All components of the NI-GAM for the 21 proxy sites along the Atlantic coast of North America. The regional component is in blue with a 95\% credible interval. The linear local component and the site-specific vertical offset are green with a 95\% credible interval. The non-linear local component is brown with a 95\% credible interval. The total posterior model fit is purple with 95\% credible interval. The x-axis is in years and y axis is in meters.}
 	\label{fig:allcomponentfulldataproxy}
 \end{figure}
 \end{sidewaysfigure}

\begin{sidewaysfigure}
 \begin{figure}[H]
 	\centering
 	\includegraphics[width=\textwidth]{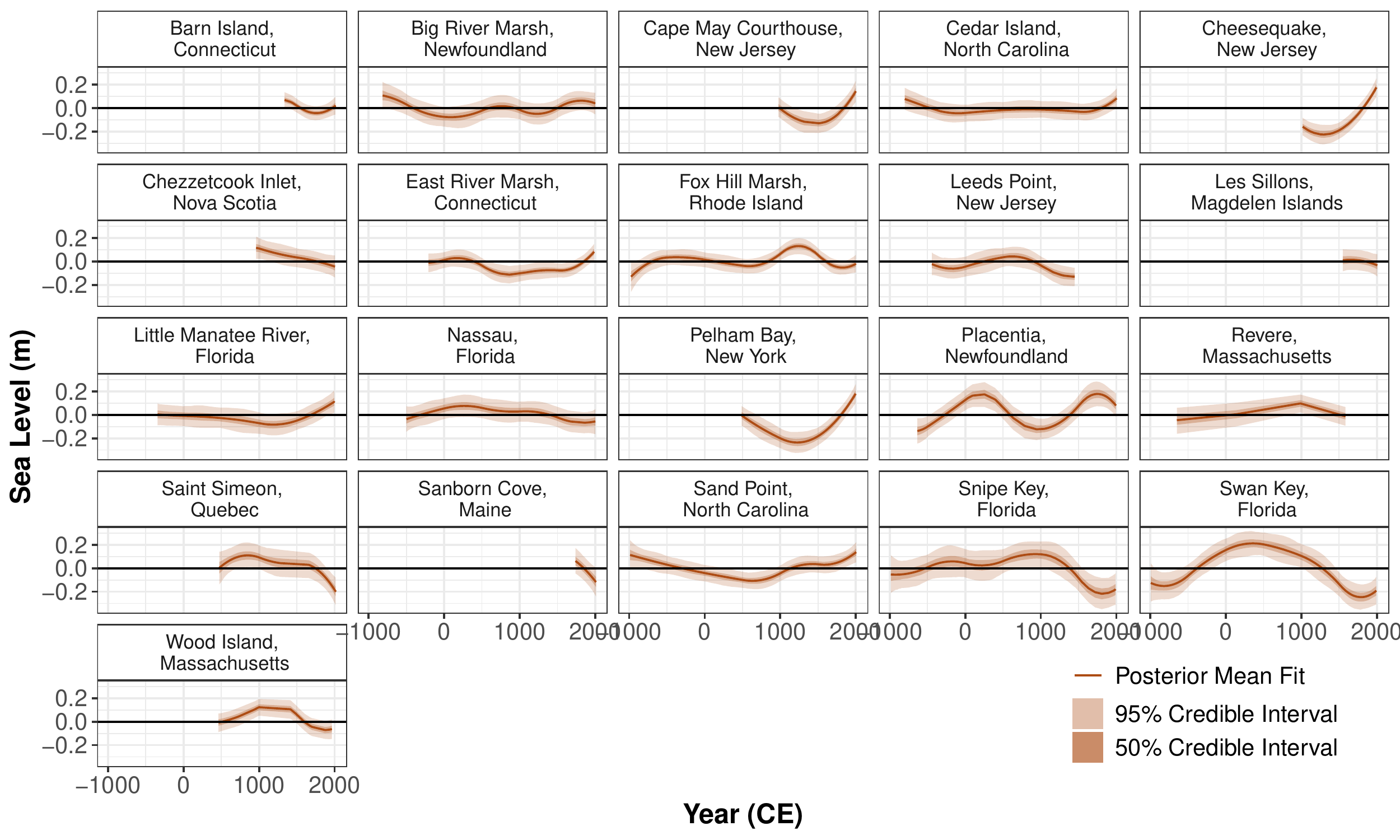}
     \caption{The non-linear local component for all proxy sites along the Atlantic coast of North America. The y-axis represents sea level in meters. The brown solid line represents the mean of the model fit with the 50 \% credible interval in dark brown shading and 95\% credible interval in the light brown shading.}
 	\label{fig:localfulldata}
 \end{figure}
\end{sidewaysfigure}

The model is run using 21 proxy sites and 66 tide gauge sites, yet we present the results of the proxy record sites as their long temporal trend provide insight into long term changes in RSL along the Atlantic coast of North America. Figure \ref{fig:totalfulldataproxy} provides the total model fit for the 21 proxy sites along the Atlantic coast of North America and Figure \ref{fig:ratefulldata} provides the rates of change for the corresponding 21 proxy sites.
Figure \ref{fig:allcomponentfulldataproxy} provides the decomposition of the NI-GAM into the total model fit and the three components; regional, linear local component with the site-specific vertical offset and non-linear local component for the 21 proxy sites.
Figure \ref{fig:regionalpostsamples} presents the regional component of the NI-GAM. The grey lines represent 10 randomly chosen posterior samples showing the underlying behaviour of the posterior for the regional component. The posterior samples are parallel resulting in reduction in the size of the uncertainty. Figure \ref{fig:localfulldata} provides the non-linear local component for each proxy record data site.

\newpage
 \subsubsection*{Model Validation}
\begin{sidewaysfigure}
 \begin{figure}[H]
 	\centering
 	\includegraphics[width=\textwidth]{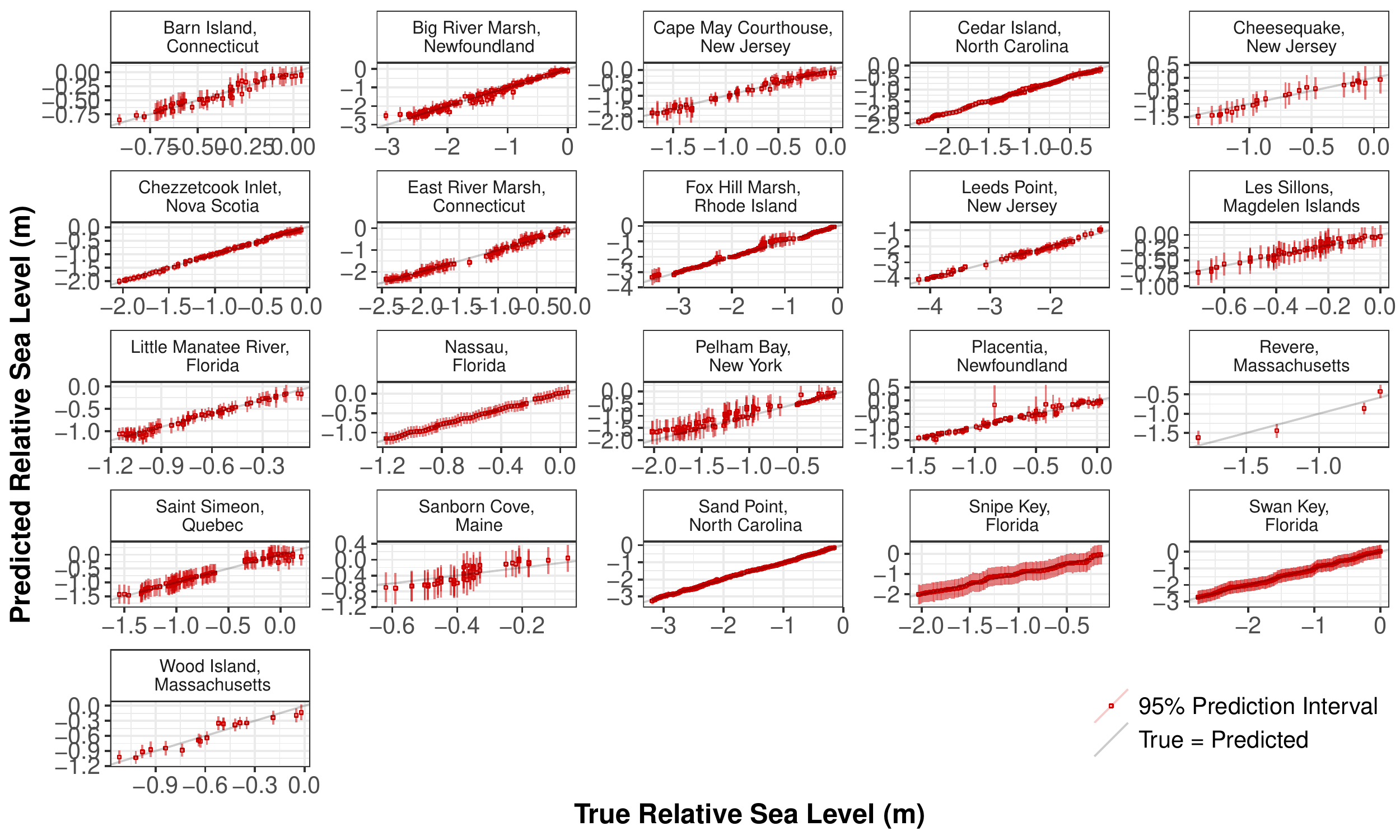}
     \caption{True versus Predicted RSL for all the 21 proxy sites along the Atlantic coast of North America using 10 fold cross validation. The grey line indicating the identity line.}
 	\label{fig:truepredfulldataprxy}
 \end{figure}
\end{sidewaysfigure}

Model validations using 10-fold cross validation was undertaken using the data set from the 21 proxy sites along the Atlantic coast of North America. The tide gauge data was not used in the 10-fold cross validation as many sites had fewer than 10 data points. Figure \ref{fig:truepredfulldataprxy} provides the true versus predicted RSL for the 21 proxy sites using 10 fold cross validation. In table \ref{Tab:coverage_all}, the empirical coverage of the model for all 21 sites along the Atlantic coast of North America is examined. The empirical coverage indicates the percentage of times the true observation lies within the prediction interval. A comparison is made between the 95\% empirical coverage and the 50\% empirical coverage. It is evident that the prediction intervals for our model are large resulting in 100\% coverage in many sites. This is due to the large size of the prediction intervals resulting from the large bivariate uncertainty that arises from the proxy records. Also included in Table \ref{Tab:coverage_all}, it the root mean square error (RMSE) for the 21 data sites along the Atlantic coast of North America which gives an insight into the prediction errors.
\begin{table}[H]
\centering
\adjustbox{max width=\textwidth}{
\begin{tabular}{lrrrrr}
  \hline
 Site Name & Empirical 95\% Coverage & 95\% PI width & Empirical 50\% Coverage & 50\% PI width & RMSE(m) \\ 
  \hline
Barn Island,
 Connecticut & 1.00 & 0.31 & 0.59 & 0.11 & 0.08 \\ 
  Big River Marsh,
 Newfoundland & 0.96 & 0.46 & 0.63 & 0.16 & 0.12 \\ 
  Cape May Courthouse,
 New Jersey & 0.99 & 0.53 & 0.81 & 0.18 & 0.14 \\ 
  Cedar Island,
 North Carolina & 1.00 & 0.26 & 0.78 & 0.09 & 0.06 \\ 
  Cheesequake,
 New Jersey & 1.00 & 0.81 & 0.78 & 0.28 & 0.21 \\ 
  Chezzetcook Inlet,
 Nova Scotia & 1.00 & 0.26 & 0.95 & 0.09 & 0.07 \\ 
  East River Marsh,
 Connecticut & 1.00 & 0.52 & 0.77 & 0.18 & 0.13 \\ 
  Fox Hill Marsh,
 Rhode Island & 0.98 & 0.40 & 0.61 & 0.14 & 0.13\\ 
  Leeds Point,
 New Jersey & 1.00 & 0.48 & 0.64 & 0.17 & 0.12\\ 
  Les Sillons,
 Magdelen Islands & 1.00 & 0.38 & 0.87 & 0.13 & 0.10 \\ 
  Little Manatee River,
 Florida & 1.00 & 0.30 & 0.84 & 0.10 & 0.07\\ 
  Nassau,
 Florida & 1.00 & 0.30 & 1.00 & 0.10 & 0.07\\ 
  Pelham Bay,
 New York & 1.00 & 0.70 & 0.51 & 0.24 & 0.19\\ 
  Placentia,
 Newfoundland & 0.97 & 0.36 & 0.59 & 0.12 & 0.11\\ 
  Revere,
 Massachusetts & 0.50 & 0.34 & 0.00 & 0.12 & 0.07 \\ 
  Saint Simeon,
 Quebec & 1.00 & 0.64 & 0.92 & 0.22 & 0.16\\ 
  Sanborn Cove,
 Maine & 1.00 & 0.72 & 0.50 & 0.25 & 0.18\\ 
  Sand Point,
 North Carolina & 1.00 & 0.33 & 0.96 & 0.11& 0.08\\ 
  Snipe Key,
 Florida & 1.00 & 0.93 & 1.00 & 0.32 & 0.23 \\ 
  Swan Key,
 Florida & 1.00 & 0.77 & 1.00 & 0.26 & 0.19\\ 
  Wood Island,
 Massachusetts & 0.78 & 0.27 & 0.28 & 0.09 & 0.06 \\ 
   \hline
\end{tabular}}
\caption{Empirical 95\% coverage for the 21 data sites along the Atlantic coast of North America with the associated prediction interval(PI). As a comparison, the prediction intervals are reduced to 50\% intervals and the empirical coverage for the 50\% is presented. The root mean square error (RMSE) is included in meters \label{Tab:coverage_all}}
\end{table}

\newpage
\section*{Acknowledgments}
Upton's and McCarthy's work is supported by A4 (Aigéin, Aeráid, agus athrú Atlantaigh) project is funded by the Marine Institute (grant: PBA/CC/18/01). Parnell’s work is supported by the SFI awards 17/CDA/4695; 16/IA/4520; 12/RC/2289P2. Kemp is supported by a U.S. National Science Foundation CAREER award (OCE-1942563). Ashe's research is funded by the U.S. National Science Foundation grant OCE-2002437 and OCE-2103754. Cahill's research is conducted with the financial support of Science Foundation Ireland and co-funded by Geological Survey Ireland under Grant number 20/FFP-P/8610.

\bibliographystyle{jmr}
\bibliography{ref.bib} 

@inbook{pugh_woodworth_2014, place={Cambridge}, title={Mean sea-level changes in time: Sea-Level Science: Understanding Tides, Surges, Tsunamis and Mean Sea-Level Changes}, DOI={10.1017/CBO9781139235778.013}, booktitle={Sea-Level Science: Understanding Tides, Surges, Tsunamis and Mean Sea-Level Changes}, publisher={Cambridge University Press}, author={Pugh, David and Woodworth, Philip}, year={2014}, pages={252–295}}

@inbook{pugh_woodworth_2014b, place={Cambridge}, title={Tidal forces: Sea-Level Science: Understanding Tides, Surges, Tsunamis and Mean Sea-Level Changes}, DOI={10.1017/CBO9781139235778.006}, booktitle={Sea-Level Science: Understanding Tides, Surges, Tsunamis and Mean Sea-Level Changes}, publisher={Cambridge University Press}, author={Pugh, David and Woodworth, Philip}, year={2014}, pages={36–59}}

@article{plummer2016rjags,
  title={{rjags: Bayesian graphical models using MCMC}},
  author={Plummer, Martyn and Stukalov, Alexey and Denwood, Matt},
  journal={R package version},
  volume={4},
  number={6},
  year={2016}
}

@article{vacchi2018,
title = {{Postglacial relative sea-level histories along the eastern Canadian coastline}},
journal = {Quaternary Science Reviews},
volume = {201},
pages = {124-146},
year = {2018},
issn = {0277-3791},
author = {Matteo Vacchi and Simon E. Engelhart and Daria Nikitina and Erica L. Ashe and W. Richard Peltier and Keven Roy and Robert E. Kopp and Benjamin P. Horton}
}

@article{shennan2018,
title = {{Relative sea-level changes and crustal movements in Britain and Ireland since the Last Glacial Maximum}},
journal = {Quaternary Science Reviews},
volume = {188},
pages = {143-159},
year = {2018},
issn = {0277-3791},
author = {Ian Shennan and Sarah L. Bradley and Robin Edwards}
}

@article{frederikse2020,
  title={The causes of sea-level rise since 1900},
  author={Frederikse, Thomas and Landerer, Felix and Caron, Lambert and Adhikari, Surendra and Parkes, David and Humphrey, Vincent W and Dangendorf, S{\"o}nke and Hogarth, Peter and Zanna, Laure and Cheng, Lijing and others},
  journal={Nature},
  volume={584},
  number={7821},
  pages={393--397},
  year={2020},
  publisher={Nature Publishing Group}
}

@article{walker2022timing,
  title={Timing of emergence of modern rates of sea-level rise by 1863},
  author={Walker, Jennifer S and Kopp, Robert E and Little, Christopher M and Horton, Benjamin P},
  journal={Nature communications},
  volume={13},
  number={1},
  pages={1--8},
  year={2022},
  publisher={Nature Publishing Group}
}

@article{plummer2003jags,
  title={{JAGS: A program for analysis of Bayesian graphical models using Gibbs sampling.}},
  author={Plummer, Martyn},
  journal={Proceedings of the 3rd International Workshop on Distributed Statistical Computing, TUWien},
  year={2003}
}

@article{Dangendorf2017,
author = {Sönke Dangendorf  and Marta Marcos  and Guy Wöppelmann  and Clinton P. Conrad  and Thomas Frederikse  and Riccardo Riva },
title = {Reassessment of 20th century global mean sea level rise},
journal = {Proceedings of the National Academy of Sciences},
volume = {114},
number = {23},
pages = {5946-5951},
year = {2017}}

@article{gelman2006prior,
  title={Prior distributions for variance parameters in hierarchical models},
  author={Gelman, Andrew},
  journal={Bayesian analysis},
  volume={1},
  number={3},
  pages={515--534},
  year={2006},
  publisher={International Society for Bayesian Analysis}
}

@article{Berrett2020,
author = {Berrett, Candace and Christensen, William F. and Sain, Stephan R. and Sandholtz, Nathan and Coats, David W. and Tebaldi, Claudia and Lopes, Hedibert F.},
title = {{Modeling sea-level processes on the U.S. Atlantic Coast}},
journal = {Environmetrics},
volume = {31},
number = {4},
pages = {e2609},
year = {2020}
}

@article{Holgate_PSMSL2013,
author = {Simon J. Holgate and Andrew Matthews and Philip L. Woodworth and Lesley J. Rickards and Mark E. Tamisiea and Elizabeth Bradshaw and Peter R. Foden and Kathleen M. Gordon and Svetlana Jevrejeva and Jeff Pugh},
title = {{New Data Systems and Products at the Permanent Service for Mean Sea Level}},
volume = {29},
journal = {Journal of Coastal Research},
number = {3},
publisher = {Coastal Education and Research Foundation},
pages = {493 -- 504},
doi = {10.2112/JCOASTRES-D-12-00175.1},
year = {2013}
}

@article{Caron2018,
author = {Caron, L. and Ivins, E. R. and Larour, E. and Adhikari, S. and Nilsson, J. and Blewitt, G.},
title = {{GIA} {M}odel {S}tatistics for {GRACE} {H}ydrology, {C}ryosphere, and {O}cean {S}cience},
journal = {Geophysical Research Letters},
volume = {45},
number = {5},
pages = {2203-2212},
year = {2018}
}

@article{Church2011,
	author = {Church, John A. and White, Neil J.},
	isbn = {1573-0956},
	journal = {Surveys in Geophysics},
	number = {4},
	pages = {585--602},
	title = {Sea-{L}evel {R}ise from the {L}ate 19th to the {E}arly 21st {C}entury},
	volume = {32},
	year = {2011}}

@article{jevrejeva2008,
  title={{Recent global sea level acceleration started over 200 years ago?}},
  author={Jevrejeva, S and Moore, John C and Grinsted, Aslak and Woodworth, Philip L},
  journal={Geophysical Research Letters},
  volume={35},
  number={8},
  year={2008},
  publisher={Wiley Online Library}
}

@article{Wenzel2010,
author = {Wenzel, Manfred and Schröter, Jens},
title = {Reconstruction of regional mean sea level anomalies from tide gauges using neural networks},
journal = {Journal of Geophysical Research: Oceans},
volume = {115},
number = {C8},
year = {2010}
}

@Article{Kopp2015,
  author={Robert Kopp and Benjamin Horton and Andrew Kemp and Claudia Tebaldi},
  title={{Past and future sea-level rise along the coast of North Carolina, USA}},
  journal={Climatic Change},
  year=2015,
  volume={132},
  number={4},
  pages={693-707},
  month={October}
}

@article{Woodworth2003,
author = {Woodworth, P.L. and Player, R.},
year = {2003},
month = {03},
pages = {287-295},
title = {{The Permanent Service for Mean Sea Level: An update to the 21st century}},
volume = {19},
journal = {Journal of Coastal Research}
}

@article{Woppelmann2006,
title = {{Tide gauges and Geodesy: a secular synergy illustrated by three present-day case studies}},
journal = {Comptes Rendus Geoscience},
volume = {338},
number = {14},
pages = {980-991},
year = {2006},
note = {La Terre observée depuis l'espace},
issn = {1631-0713},
doi = {https://doi.org/10.1016/j.crte.2006.07.006},
author = {Guy Wöppelmann and Susanna Zerbini and Marta Marcos},
}

@article{Sachs1977,
  title={Paleoecological transfer functions},
  author={Sachs, Harvey Maurice and Webb III, T and Clark, DR},
  journal={Annual Review of Earth and Planetary Sciences},
  volume={5},
  number={1},
  pages={159--178},
  year={1977},
  publisher={Annual Reviews 4139 El Camino Way, PO Box 10139, Palo Alto, CA 94303-0139, USA}
}

@article{Redfield1972,
  title={{Development of a New England salt marsh}},
  author={Redfield, Alfred C},
  journal={Ecological monographs},
  volume={42},
  number={2},
  pages={201--237},
  year={1972},
  publisher={Wiley Online Library}
}

@article{Donnelly2004,
  title={Coupling instrumental and geological records of sea-level change: Evidence from southern {N}ew {E}ngland of an increase in the rate of sea-level rise in the late 19th century},
  author={Donnelly, Jeffrey P and Cleary, Peter and Newby, Paige and Ettinger, Robert},
  journal={Geophysical Research Letters},
  volume={31},
  number={5},
  year={2004},
  publisher={Wiley Online Library}
}

@article{Cahill2016,
author = {Cahill, N. and Kemp, A. C. and Horton, B. P. and Parnell, A. C.},
title = {A {B}ayesian hierarchical model for reconstructing relative sea level: from raw data to rates of change},
journal = {Climate of the Past},
volume = {12},
year = {2016},
number = {2},
pages = {525--542}
}

@article{Horton2006,
  title={{Quantifying Holocene sea level change using intertidal foraminifera: lessons from the British Isles}},
  author={Horton, Benjamin P and Edwards, Robin J},
  journal={Departmental Papers (EES)},
  pages={50},
  year={2006}
}

@article {Kemp2011,
	author = {Kemp, Andrew C. and Horton, Benjamin P. and Donnelly, Jeffrey P. and Mann, Michael E. and Vermeer, Martin and Rahmstorf, Stefan},
	title = {Climate related sea-level variations over the past two millennia},
	volume = {108},
	number = {27},
	pages = {11017--11022},
	year = {2011},
	doi = {10.1073/pnas.1015619108},
	publisher = {National Academy of Sciences},
	journal = {Proceedings of the National Academy of Sciences}
}

@article{Walker2021,
	author = {Walker, Jennifer S. and Kopp, Robert E. and Shaw, Timothy A. and Cahill, Niamh and Khan, Nicole S. and Barber, Donald C. and Ashe, Erica L. and Brain, Matthew J. and Clear, Jennifer L. and Corbett, D. Reide and Horton, Benjamin P.},
	doi = {10.1038/s41467-021-22079-2},
	isbn = {2041-1723},
	journal = {Nature Communications},
	number = {1},
	pages = {1841},
	title = {{Common Era sea-level budgets along the U.S. Atlantic coast}},
	volume = {12},
	year = {2021},
}

@article{Gregory2019ConceptsGlobal,
    title = {{Concepts and Terminology for Sea Level: Mean, Variability and Change, Both Local and Global}},
    year = {2019},
    journal = {Surveys in Geophysics},
    author = {Gregory, Jonathan M. and Griffies, Stephen M. and Hughes, Chris W. and Lowe, Jason A. and Church, John A. and Fukimori, Ichiro and Gomez, Natalya and Kopp, Robert E. and Landerer, Felix and Cozannet, Gonéri Le and Ponte, Rui M. and Stammer, Detlef and Tamisiea, Mark E. and van de Wal, Roderik S.W.},
    number = {6},
    month = {11},
    pages = {1251--1289},
    volume = {40},
    publisher = {Springer Netherlands},
    issn = {15730956}
}

@article{Kemp2013Sea-levelUSA,
    title = {{Sea-level change during the last 2500 years in New Jersey, USA}},
    year = {2013},
    journal = {Quaternary Science Reviews},
    author = {Kemp, Andrew C. and Horton, Benjamin P. and Vane, Christopher H. and Bernhardt, Christopher E. and Corbett, D. Reide and Engelhart, Simon E. and Anisfeld, Shimon C. and Parnell, Andrew C. and Cahill, Niamh},
    pages = {90--104},
    volume = {81},
    doi = {10.1016/j.quascirev.2013.09.024},
    issn = {02773791}
}

@article{Gehrels1994,
 author = {W. Roland Gehrels},
 journal = {Journal of Coastal Research},
 number = {4},
 pages = {990--1009},
 publisher = {Coastal Education \& Research Foundation, Inc.},
 title = {{Determining Relative Sea-Level Change from Salt-Marsh Foraminifera and Plant Zones on the Coast of Maine, U.S.A.}},
 volume = {10},
 year = {1994}
}

@article{Khan2017,
title = {{Drivers of Holocene sea-level change in the Caribbean}},
journal = {Quaternary Science Reviews},
volume = {155},
pages = {13-36},
year = {2017},
issn = {0277-3791},
doi = {https://doi.org/10.1016/j.quascirev.2016.08.032},
author = {Nicole S. Khan and Erica Ashe and Benjamin P. Horton and Andrea Dutton and Robert E. Kopp and Gilles Brocard and Simon E. Engelhart and David F. Hill and W.R. Peltier and Christopher H. Vane and Fred N. Scatena}
}

@article{Stammer2013,
author = {Stammer, Detlef and Cazenave, Anny and Ponte, Rui M. and Tamisiea, Mark E.},
title = {{Causes for Contemporary Regional Sea Level Changes}},
journal = {Annual Review of Marine Science},
volume = {5},
number = {1},
pages = {21-46},
year = {2013}}

@article{2K_2019,
	author = {Neukom, Raphael and Barboza, Luis A. and Erb, Michael P. and Shi, Feng and Emile-Geay, Julien and Evans, Michael N. and Franke, J{\"o}rg and Kaufman, Darrell S. and L{\"u}cke, Lucie and Rehfeld, Kira and Schurer, Andrew and Zhu, Feng and Br{\"o}nnimann, Stefan and Hakim, Gregory J. and Henley, Benjamin J. and Ljungqvist, Fredrik Charpentier and McKay, Nicholas and Valler, Veronika and von Gunten, Lucien and PAGES 2k Consortium},
	journal = {Nature Geoscience},
	number = {8},
	pages = {643--649},
	title = {Consistent multidecadal variability in global temperature reconstructions and simulations over the Common Era},
	volume = {12},
	year = {2019}}

@article{Shennan2015_Handbook,
author = {Shennan, Ian and Long, Antony J. and Horton, Benjamin P.},
journal = {John Wiley \& Sons, Ltd},
isbn = {9781118452547},
title = {Handbook of Sea‐Level Research},
year = {2015}}

@inbook{Parnell2015_Handbook,
author = {Parnell, Andrew C. and Gehrels, W. Roland},
publisher = {John Wiley \& Sons, Ltd},
title = {Using chronological models in late Holocene sea-level reconstructions from saltmarsh sediments},
booktitle = {Handbook of Sea‐Level Research},
chapter = {32},
pages = {500-513},
year = {2015}
}

@article{Meltzner2017,
	author = {Meltzner, Aron J. and Switzer, Adam D. and Horton, Benjamin P. and Ashe, Erica and Qiu, Qiang and Hill, David F. and Bradley, Sarah L. and Kopp, Robert E. and Hill, Emma M. and Majewski, J{\c{e}}drzej M. and Natawidjaja, Danny H. and Suwargadi, Bambang W.},
	doi = {10.1038/ncomms14387},
	isbn = {2041-1723},
	journal = {Nature Communications},
	number = {1},
	pages = {14387},
	title = {{Half-metre sea-level fluctuations on centennial timescales from mid-Holocene corals of Southeast Asia}},
	volume = {8},
	year = {2017}}

@article{Grinsted2015,
    title = {{Projected Change—Sea Level}},
    journal = {Second assessment of climate change for Baltic Sea basin},
    year = {2015},
    author = {Grinsted, Aslak},
    chapter = {14},
    pages = {253--263},
    publisher = {Springer, Cham},
}

@book{Dey2000,
 author={Dey, Dipak K and Ghosh, Sujit K and Mallick, Bani K},
 title = {Generalized linear models: a {B}ayesian perspective},
publisher = {CRC Press},
year = {2000}
}

@article{Horton2018MappingProbability,
    title = {{Mapping Sea-Level Change in Time, Space, and Probability}},
    year = {2018},
    journal = {Annual Review of Environment and Resources},
    author = {Horton, Benjamin P. and Kopp, Robert E. and Garner, Andra J. and Hay, Carling C. and Khan, Nicole S. and Roy, Keven and Shaw, Timothy A.},
    number = {1},
    pages = {481--521},
    volume = {43},
    doi = {10.1146/annurev-environ-102017-025826},
    issn = {1543-5938}
}

@article{Whitehouse2018,
    title = {{Glacial isostatic adjustment modelling: historical perspectives, recent advances, and future directions}},
    year = {2018},
    journal = {Earth Surf. Dynam},
    author = {Whitehouse, Pippa L},
    pages = {401--429},
    volume = {6}
}

@article{Peltier2004,
    author = {Peltier, W.R},
    year = {2004},
    title = {{Global Glacial Isostasy and the Surface of the Ice-Age Earth: The ICE-5G (VM2) Model and GRACE}},
    journal = {Annual Review of Earth and Planetary Sciences},
    volume = {32},
    pages = {111--149}
    }

@article{Cahill2015aStats,
  author = {Cahill, Niamh and Kemp, Andrew C and Horton, Benjamin P and Parnell, Andrew C},
  doi = {10.1214/15-AOAS824},
  journal = {The Annals of Applied Statistics},
  number = {2},
  pages = {547--571},
  title = {{Modeling sea-level change using {E}rrors-in-{V}ariables integrated {G}aussian {P}rocess 1}},
  volume = {9},
  year = {2015}
}

@article{Donnelly2006,
 author = {Jeffrey P. Donnelly},
 journal = {Journal of Coastal Research},
 number = {5},
 pages = {1051--1061},
 publisher = {Coastal Education \& Research Foundation, Inc.},
 title = {A {R}evised {L}ate {H}olocene {S}ea-{L}evel {R}ecord for {N}orthern {M}assachusetts, {USA}},
 volume = {22},
 year = {2006}
}

@article{Gerlach2017,
title = {{Reconstructing Common Era relative sea-level change on the Gulf Coast of Florida}},
journal = {Marine Geology},
volume = {390},
pages = {254-269},
year = {2017},
issn = {0025-3227},
doi = {https://doi.org/10.1016/j.margeo.2017.07.001},
author = {Matthew J. Gerlach and Simon E. Engelhart and Andrew C. Kemp and Ryan P. Moyer and Joseph M. Smoak and Christopher E. Bernhardt and Niamh Cahill}
}

@article{Barnett2019,
title = {Late {H}olocene sea-level changes in eastern {Q}uébec and potential drivers},
journal = {Quaternary Science Reviews},
author = {R.L. Barnett and P. Bernatchez and M. Garneau and M.J. Brain and D.J. Charman and D.B. Stephenson and S. Haley and N. Sanderson},
volume = {203},
pages = {151-169},
year = {2019},
issn = {0277-3791},
doi = {https://doi.org/10.1016/j.quascirev.2018.10.039}
}

@article{Gehrels2020,
author = {Gehrels, W. R. and Dangendorf, S. and Barlow, N. L. M. and Saher, M. H. and Long, A. J. and Woodworth, P. L. and Piecuch, C. G. and Berk, K.},
title = {{A Preindustrial Sea-Level Rise Hotspot Along the Atlantic Coast of North America}},
journal = {Geophysical Research Letters},
volume = {47},
number = {4},
doi = {https://doi.org/10.1029/2019GL085814},
year = {2020}
}

@article{Kemp2015,
title = {{Relative sea-level change in Connecticut (USA) during the last 2200 yrs}},
journal = {Earth and Planetary Science Letters},
volume = {428},
pages = {217-229},
year = {2015},
issn = {0012-821X},
doi = {https://doi.org/10.1016/j.epsl.2015.07.034},
author = {Andrew C. Kemp and Andrea D. Hawkes and Jeffrey P. Donnelly and Christopher H. Vane and Benjamin P. Horton and Troy D. Hill and Shimon C. Anisfeld and Andrew C. Parnell and Niamh Cahill}
}

@article{Roy2015,
    author = {Roy, Keven and Peltier, W.R.},
    title = {"{Glacial isostatic adjustment, relative sea level history and mantle viscosity: reconciling relative sea level model predictions for the U.S. East coast with geological constraints}"},
    journal = {Geophysical Journal International},
    volume = {201},
    number = {2},
    pages = {1156-1181},
    year = {2015}

}

@article{Khan2022,
title = {{Relative sea-level change in South Florida during the past ~5000 years}},
journal = {Global and Planetary Change},
volume = {216},
pages = {103902},
year = {2022},
issn = {0921-8181},
doi = {https://doi.org/10.1016/j.gloplacha.2022.103902},
author = {Nicole S. Khan and Erica Ashe and Ryan P. Moyer and Andrew C. Kemp and Simon E. Engelhart and Matthew J. Brain and Lauren T. Toth and Amanda Chappel and Margaret Christie and Robert E. Kopp and Benjamin P. Horton}
}

@inbook{Kemp2015_SLhandbook,
author = {Kemp, Andrew C. and Telford, Richard J.},
publisher = {John Wiley \& Sons, Ltd},
isbn = {9781118452547},
title = {Transfer functions: Handbook of Sea‐Level Research},
booktitle = {Handbook of Sea‐Level Research},
chapter = {31},
pages = {470-499},
doi = {https://doi.org/10.1002/9781118452547.ch31},
year = {2015}

}

@inbook{Edwards2015_SLhandbook,
author = {Edwards, Robin and Wright, Alex},
publisher = {John Wiley \& Sons, Ltd},
isbn = {9781118452547},
title = {Foraminifera: Handbook of Sea‐Level Research},
chapter = {13},
pages = {191-217},
doi = {https://doi.org/10.1002/9781118452547.ch13},
year = {2015}
}

@inbook{Marshall2015_SLhandbook,
author = {Marshall, Wil},
publisher = {John Wiley \& Sons, Ltd},
isbn = {9781118452547},
title = {Chronohorizons: Handbook of Sea‐Level Research},
booktitle = {Handbook of Sea‐Level Research},
chapter = {25},
pages = {373-385},
doi = {https://doi.org/10.1002/9781118452547.ch25},
year = {2015}
}

@inbook{Torn2015SL_handbook,
author = {Törnqvist, Torbjörn E. and Rosenheim, Brad E. and Hu, Ping and Fernandez, Alvaro B.},
publisher = {John Wiley \& Sons, Ltd},
isbn = {9781118452547},
title = {Radiocarbon dating and calibration: Handbook of Sea‐Level Research},
booktitle = {Handbook of Sea‐Level Research},
chapter = {23},
pages = {347-360},
doi = {https://doi.org/10.1002/9781118452547.ch23},
year = {2015}
}

@article{Wood2006a,
author = {Wood, Simon N.},
title = {{Low-Rank Scale-Invariant Tensor Product Smooths for Generalized Additive Mixed Models}},
journal = {Biometrics},
volume = {62},
number = {4},
pages = {1025-1036},
year = {2006}
}

@misc{Stearns2017,
	author = {Stearns, Rachel B. and Engelhart, Simon E.},
	year = {2017},
	db = {ProQuest Dissertations \& Theses A\&I; ProQuest One Academic},
	isbn = {978-0-355-13981-5},
	journal = {ProQuest Dissertations and Theses},
	number = {10615081},
	pages = {87},
	pp = {United States -- Rhode Island},
	publisher = {University of Rhode Island},
	title = {{A High-Resolution Reconstruction of Late-Holocene Relative Sea Level in Rhode Island, USA}}}

@article{Barnett2017,
author = {Barnett, Robert L. and Bernatchez, Pascal and Garneau, Michelle and Juneau, Marie-Noëlle},
title = {Reconstructing late {H}olocene relative sea-level changes at the {M}agdalen {I}slands ({G}ulf of {S}t. {L}awrence, {C}anada) using multi-proxy analyses},
journal = {Journal of Quaternary Science},
volume = {32},
number = {3},
pages = {380-395},
year = {2017}
}

@article{Kemp2014,
title = {{Late Holocene sea- and land-level change on the U.S. southeastern Atlantic coast}},
journal = {Marine Geology},
volume = {357},
pages = {90-100},
year = {2014},
issn = {0025-3227},
doi = {https://doi.org/10.1016/j.margeo.2014.07.010},
author = {Andrew C. Kemp and Christopher E. Bernhardt and Benjamin P. Horton and Robert E. Kopp and Christopher H. Vane and W. Richard Peltier and Andrea D. Hawkes and Jeffrey P. Donnelly and Andrew C. Parnell and Niamh Cahill}
}

@book{dierckx1995curve,
  title={Curve and surface fitting with splines},
  author={Dierckx, Paul},
  year={1995},
  publisher={Oxford University Press}
}

@article{Peltier2014,
author = {Argus, Donald and Peltier, W. and Drummond, Rosemarie and Moore, Angelyn},
year = {2014},
month = {05},
pages = {537-563},
title = {The {A}ntarctica component of postglacial rebound model {ICE-6GC} ({VM}5a) based on {GPS} positioning, exposure age dating of ice thicknesses, and relative sea level histories},
volume = {198},
journal = {Geophysical Journal International},
doi = {10.1093/gji/ggu140}
}

@article{deBoor1978,
author = {de Boor, Carl},
year = {1978},
month = {01},
title = {A {P}ractical {G}uide to {S}pline},
volume = {27},
journal = {Applied Mathematical Sciences, New York: Springer, 1978},
doi = {10.2307/2006241}
}

@article{Kemp2017,
author = {Andrew C Kemp and Troy D Hill and Christopher H Vane and Niamh Cahill and Philip M Orton and Stefan A Talke and Andrew C Parnell and Kelsey Sanborn and Ellen K Hartig},
title ={{Relative sea-level trends in New York City during the past 1500 years}},
journal = {The Holocene},
volume = {27},
number = {8},
pages = {1169-1186},
year = {2017},
doi = {10.1177/0959683616683263}
}

@article{Engelhart2009,
    author = {Engelhart, Simon E. and Horton, Benjamin P. and Douglas, Bruce C. and Peltier, W. Richard and Törnqvist, Torbjörn E.},
    title = "{{Spatial variability of late Holocene and 20th century sea-level rise along the Atlantic coast of the United States}}",
    journal = {Geology},
    volume = {37},
    number = {12},
    pages = {1115-1118},
    year = {2009},
    month = {12},
    issn = {0091-7613},
    doi = {10.1130/G30360A.1}
}

@article{Blaauw2011BACON,
author = {Maarten Blaauw and J. Andr{\'e}s Christen},
title = {{Flexible paleoclimate age-depth models using an autoregressive gamma process}},
volume = {6},
journal = {Bayesian Analysis},
number = {3},
publisher = {International Society for Bayesian Analysis},
pages = {457 -- 474},
year = {2011},
doi = {10.1214/11-BA618}
}

@article{Aquino2018_Rplum,
	author = {Aquino-L{\'o}pez, Marco A. and Blaauw, Maarten and Christen, J. Andr{\'e}s and Sanderson, Nicole K.},
	journal = {Journal of Agricultural, Biological and Environmental Statistics},
	number = {3},
	pages = {317--333},
	title = {Bayesian {A}nalysis of 210 {Pb} {D}ating},
	volume = {23},
	year = {2018}}

@article{WRIGHT2017,
title = {Reconstructing the accumulation history of a saltmarsh sediment core: {W}hich age-depth model is best?},
journal = {Quaternary Geochronology},
volume = {39},
pages = {35-67},
year = {2017},
issn = {1871-1014},
doi = {https://doi.org/10.1016/j.quageo.2017.02.004},
author = {Alexander J. Wright and Robin J. Edwards and Orson {van de Plassche} and Maarten Blaauw and Andrew C. Parnell and Klaas {van der Borg} and Arie F.M. {de Jong} and Helen M. Roe and Katherine Selby and Stuart Black}
}

@misc{Gabry2017shinystan,
    author = {Gabry, Jonah and Goodrich, Ben},
    year = {2017},
    title = {{rstanarm: Bayesian applied regression modeling via Stan}},
    journal = {R package version 2.15.3}}

@article{Porcu2021,
    title = {{30 Years of space-time covariance functions}},
    author = {Porcu, Emilio and Furrer, Reinhard and Nychka, Douglas}, 
    journal = {WIREs Computational Statistics},
    volume = {13},
    number = {2},
    pages = {1512},
    year = {2021}}

@ARTICLE{Piecuch2017,
	author = {Piecuch, Christopher G. and Huybers, Peter and Tingley, Martin P.},
	title = {Comparison of full and empirical Bayes approaches for inferring sea-level changes from tide-gauge data},
	year = {2017},
	journal = {Journal of Geophysical Research: Oceans},
	volume = {122},
	number = {3},
	pages = {2243 – 2258}
}

@article{Hay2015,
author = {Hay, Carling C. and Morrow, Eric and Kopp, Robert E. and Mitrovica, Jerry X.},
journal = {Nature},
number = {7535},
pages = {481--484},
publisher = {Nature Publishing Group},
title = {{Probabilistic reanalysis of twentieth-century sea-level rise}},
volume = {517},
year = {2015}}

@article{Plummer2015,
	author = {Plummer, Martyn},
	journal = {Statistics and Computing},
	number = {1},
	pages = {37--43},
	title = {{Cuts in Bayesian graphical models}},
	volume = {25},
	year = {2015}}

@article{Kopp2016,
  author = {Kopp, Robert E. and Kemp, Andrew C. and Bittermann, Klaus and Horton, Benjamin P. and Donnelly, Jeffrey P. and Gehrels, W. Roland and Hay, Carling C. and Mitrovica, Jerry X. and Morrow, Eric D. and Rahmstorf, Stefan},
  doi = {10.1073/pnas.1517056113},
  issn = {10916490},
  journal = {Proceedings of the National Academy of Sciences of the United States of America},
  number = {11},
  pages = {E1434--E1441},
  title = {{Temperature-driven global sea-level variability in the Common Era}},
  volume = {113},
  year = {2016}
}

@article{Ashe2019,
  author = {Ashe, Erica L. and Cahill, Niamh and Hay, Carling and Khan, Nicole S. and Kemp, Andrew and Engelhart, Simon E. and Horton, Benjamin P. and Parnell, Andrew C. and Kopp, Robert E.},
  journal = {Quaternary Science Reviews},
  pages = {58--77},
  publisher = {Elsevier Ltd},
  title = {{Statistical modeling of rates and trends in Holocene relative sea level}},
  volume = {204},
  year = {2019}
}

@article{parnell2008,
  title={A flexible approach to assessing synchroneity of past events using Bayesian reconstructions of sedimentation history},
  author={Parnell, Andrew C and Haslett, John and Allen, Judy RM and Buck, Caitlin E and Huntley, Brian},
  journal={Quaternary Science Reviews},
  volume={27},
  number={19-20},
  pages={1872--1885},
  year={2008},
  publisher={Elsevier}
}

@article{McHutchon2011,
  author = {McHutchon, Andrew and Rasmussen, Carl Edward},
  isbn = {9781618395993},
  journal = {Advances in Neural Information Processing Systems 24: 25th Annual Conference on Neural Information Processing Systems 2011, NIPS 2011},
  pages = {1--9},
  title = {{Gaussian Process training with input noise}},
  year = {2011}
}

@book{wood_2017, 
  title={{Generalized additive models: an introduction with R}}, 
  publisher={CRC Press}, 
  author={Wood, Simon N.}, 
  year={2017}
}

@misc{Church2001,
author = {J. A. Church and J. M. Gregory and Philippe Huybrechts and M. Kuhn and K. Lambeck and M. T. Nhuan and D. Qin and P. L. Woodworth},
         journal = {IPCC 2001},
title = {Changes in {S}ea {L}evel: Climate Change 2001: The {S}cientific {B}asis. {C}ontribution of {W}orking {G}roup {I} to the {T}hird {A}ssessment {R}eport of the {I}ntergovernmental {P}anel},
         pages = {639--694},
         year = {2001}
}

@article{eilers_1996,
  author = {Eilers, Paul and Marx, Brian},
  journal = {Statistical Science},
  volume = {11},
  title = {{Flexible Smoothing with B-splines and Penalities}},
  year = {1996}
}

@article{IPCC_2013_chp13,
  author = {Church, John A. and Clark, Peter U.},
  journal = {IPCC 2013},
  title = {Climate {C}hange 2013: The {P}hysical {S}cience {B}asis. {C}ontribution of {W}orking {G}roup {I} to the {F}ifth {A}ssessment {R}eport of the {I}ntergovernmental {P}anel on {C}limate {C}hange},
  volume = {13},
  publisher = {Cambridge University Press},
  year = {2013}}

@article{Spiegel2002,
author = {Spiegelhalter, David J. and Best, Nicola G. and Carlin, Bradley P. and Van Der Linde, Angelika},
title = {Bayesian measures of model complexity and fit},
journal = {Journal of the Royal Statistical Society: Series B (Statistical Methodology)},
volume = {64},
number = {4},
pages = {583-639},
year = {2002}}

@article{plummer2006coda,
  title={{CODA: convergence diagnosis and output analysis for MCMC}},
  author={Plummer, Martyn and Best, Nicky and Cowles, Kate and Vines, Karen},
  journal={R news},
  volume={6},
  number={1},
  pages={7--11},
  year={2006}
}

@article{IPCC2021summary,
author = {Masson-Delmotte, Valérie and Zhai, Panmao and Pirani, Anna and Connors, Sarah L. and Péan, Clotilde and Berger, Sophie and Caud, Nada and Chen, Yang and Goldfarb, Leah and Gomis, Melissa I. and Huang, Mengtian and Leitzell, Katherine and Lonnoy, Elisabeth and Matthews, J.B. Robin and Maycock, Thomas K. and Waterfield, Tim and Yelekçi, Ozge and Yu, Rong and Zhou, Baiquan},
journal = {Cambridge University Press},
title = {{Summary for Policymakers. In: Climate Change 2021: The Physical Science Basis. Contribution of
Working Group I to the Sixth Assessment Report of the Intergovernmental Panel on Climate Change}},
year = {2021}
}

@article{Kemp2018,
  title={{Relative sea-level change in Newfoundland, Canada during the past~ 3000 years}},
  author={Kemp, Andrew C and Wright, Alexander J and Edwards, Robin J and Barnett, Robert L and Brain, Matthew J and Kopp, Robert E and Cahill, Niamh and Horton, Benjamin P and Charman, Dan J and Hawkes, Andrea D and others},
  journal={Quaternary Science Reviews},
  volume={201},
  pages={89--110},
  year={2018},
  publisher={Elsevier}
}

\end{document}